\documentclass[%
 reprint,
superscriptaddress,
nofootinbib,
 amsmath,amssymb,
 aps,
prx,
nolongbibliography,
]{revtex4-2}
\usepackage[english]{babel}
\usepackage[colorlinks=true,linkcolor=blue,citecolor=blue,urlcolor=blue]{hyperref}
\usepackage{graphicx}
\usepackage{placeins}
\usepackage{dcolumn}
\usepackage{bm}
\usepackage{float}

\usepackage{comment}

\newcommand{\ket}{\rangle}
\newcommand{\bra}{\langle}

\newcommand{\degree}{$^{\circ}$}

\usepackage{rotating}

\usepackage[dvipsnames]{xcolor}

\begin{document}

\title{Multivalent optical cycling centers in polyatomic molecules}

\author{Phelan Yu}
\email{phelanyu@caltech.edu}

\affiliation{Division of Physics, Mathematics, and Astronomy, California Institute of Technology, Pasadena, California 91125, USA}
\affiliation{Institute for Quantum Information and Matter, California Institute of Technology, Pasadena, California 91125, USA}

\author{Adrian Lopez}
\affiliation{Division of Physics, Mathematics, and Astronomy, California Institute of Technology, Pasadena, California 91125, USA}
\author{William A. Goddard III}
\affiliation{Materials and Process Simulation Center, California Institute of Technology, Pasadena, California 91125, USA}
\affiliation{Division of Chemistry and Chemical Engineering, California Institute of Technology, Pasadena, California 91125, USA}
\author{Nicholas R. Hutzler}
\affiliation{Division of Physics, Mathematics, and Astronomy, California Institute of Technology, Pasadena, California 91125, USA}

\date{\today}

\begin{abstract}
Optical control of polyatomic molecules promises new opportunities in precision metrology, fundamental chemistry, quantum information, and many-body science. Contemporary experimental and theoretical efforts have mostly focused on cycling photons via excitation of a single electron localized to an alkaline earth (group 2)-like metal center. In this manuscript, we consider pathways towards optical cycling in polyatomic molecules with multi-electron degrees of freedom, which arise from two or more cycling electrons localized to $p$-block post-transition metal and metalloid (group 13, 14, and 15) centers. We characterize the electronic structure and rovibrational branching of several prototypical candidates using \textit{ab initio} quantum chemical methods. Despite increased internal complexity and challenging design parameters, we find several molecules possessing quasi-closed photon cycling schemes with highly diagonal, visible and near-infrared transitions. Furthermore, we identify new heuristics for engineering optically controllable and laser-coolable polyatomic molecules with multi-electron cycling centers. Our results help elucidate the interplay between hybridization, repulsion, and ionicity in optically active species and provide a first step towards using polyatomic molecules with complex electronic structure as a resource for quantum science and measurement.
\end{abstract}

\maketitle
\section{Introduction}\label{sec:intro}
Cold molecules are powerful platforms for exploring a range of fundamental questions in physics and chemistry. Unique mechanical, spin, and dipolar degrees of freedom available in molecules enable new possibilities in quantum information \cite{demille_quantum_2002,ni_dipolar_2018,yu_scalable_2019, albert_robust_2020} and simulation \cite{moses_new_2017,bohn_cold_2017,blackmore_ultracold_2018}, precision measurement and metrology \cite{kozyryev_precision_2017,hutzler_polyatomic_2020}, as well as state-resolved chemistry \cite{balakrishnan_perspective:_2016,puri_synthesis_2017,hu_direct_2019}. In the last five years, laser cooling and optical control have been extended to increasingly complex polyatomic molecules, paving the way towards the high phase space density \cite{cheuk_-enhanced_2018, ding_sub-doppler_2020} and coherent quantum control \cite{anderegg_optical_2019,cheuk_observation_2020,caldwell_long_2020} necessary for realizing science applications with cold gases of optically active polyatomic molecules. Simultaneously, theoretical understanding of the features that make molecules amenable to optical cycling and laser cooling has significantly advanced \cite{isaev_polyatomic_2016, kozyryev_proposal_2016, orourke_hypermetallic_2019,ivanov_two_2020, ivanov_towards_2019, li_emulating_2019, ivanov_search_2020, klos_prospects_2020,  augenbraun_molecular_2020, dickerson_franck-condon_2021,mitra_pathway_2022,zhu_functionalizing_2022}, leading to the identification of several classes of polyatomics with favorable chemical and structural configurations. 

A key characteristic of photon cycling in  molecules is the presence of valence electrons localized to metallic optical cycling centers (OCCs), which enable rapid, repeated scattering of resonant photons for optical state control, detection, and cooling. The simplest ``monovalent" OCCs can be engineered by bonding an alkaline earth-like (AEL) metal\footnote{This includes alkaline earths (Be, Mg, Ca, Sr, Ba, Ra) and transition metals with $s^2$ valence and filled $d$/$f$-shells (e.g. Yb, Hg)} to a one-electron acceptor or pseudohalogen ligand \cite{isaev_polyatomic_2016, kozyryev_proposal_2016, tarbutt_laser_2019, mccarron_laser_2018, fitch_laser-cooled_2021}, forming an open-shell molecule (e.g. SrF \cite{barry_magneto-optical_2014,mccarron_improved_2015,steinecker_improved_2016,norrgard_submillikelvin_2016,  langin_polarization_2021}, CaF \cite{zhelyazkova_laser_2014, truppe_molecules_2017, anderegg_laser_2018, cheuk_-enhanced_2018, anderegg_optical_2019}, YbF \cite{lim_laser_2018, alauze_ultracold_2021}, BaH \cite{mcnally_optical_2020}, BaF \cite{albrecht_buffer-gas_2020, chen_radiative_2017}) with an excited electronic structure roughly similar to alkali atoms. The remaining $s\sigma$ electron on the metal is then polarized away from the ionic metal-ligand bond. Metal-centered, atom-like electronic excitations are highly decoupled from the rovibrational modes of the molecule, with only a handful of repumping lasers needed to scatter $10^3-10^5$ photons \cite{di_rosa_laser-cooling_2004,isaev_polyatomic_2016,baum_establishing_2021,stuhl_magneto-optical_2008, augenbraun_molecular_2020}. This heuristic has been very successful at identifying laser-coolable molecules, and all polyatomic species laser cooled to date (SrOH \cite{kozyryev_sisyphus_2017}, CaOH \cite{baum_1d_2020, vilas_magneto-optical_2021}, YbOH \cite{augenbraun_laser-cooled_2020}, CaOCH$_3$ \cite{mitra_direct_2020}) have followed the AEL-pseudohalogen template to form single electron, alkali-like OCCs.

A natural question then follows: is it possible to design molecules containing optical centers with \textit{multiple} localized cycling electrons, while preserving key structural features that enable optical control and laser cooling? In cold atom experiments, multi-electron degrees of freedom provide versatile mechanisms for both controlling and studying the behavior of complex quantum systems. Individual atoms that possess two (or more) valence cycling electrons, such as AEL atoms, give rise to electronic states with metastable lifetimes \cite{yasuda_lifetime_2004, jensen_experimental_2011}, ultranarrow optical transitions \cite{katori_magneto-optical_1999,kuwamoto_magneto-optical_1999, binnewies_doppler_2001},  perturbation-free ``magic" trapping conditions \cite{cooper_alkaline_2018, norcia_microscopic_2018, saskin_narrow-line_2019}, efficient autoionization pathways \cite{cooke_doubly_1978,lochead_number-resolved_2013,madjarov_high-fidelity_2020}, and fully tunable couplings to internal spins \cite{boyd_nuclear_2007}. 

Leveraging these features in multi-electron atoms has been a principal factor enabling record-setting optical lattice \cite{ushijima_cryogenic_2015,campbell_fermi-degenerate_2017, takamoto_test_2020, bothwell_resolving_2022} and tweezer clocks \cite{madjarov_atomic-array_2019, young_half-minute-scale_2020}, analog many-body simulators of SU($N$) and multiorbital Hamiltonians \cite{gorshkov_two-orbital_2010,taie_su6_2012,hofrichter_direct_2016,ozawa_antiferromagnetic_2018,sonderhouse_thermodynamics_2020, schafer_tools_2020}, advanced atom interferometers \cite{rudolph_large_2020}, high-fidelity entangling gates \cite{madjarov_high-fidelity_2020, pagano_fast_2019,daley_quantum_2008}, and telecom-compatible quantum transducers and memories \cite{covey_telecom-band_2019, gorshkov_alkaline-earth-metal_2009, jenkins_ytterbium_2022, ma_universal_2022, barnes_assembly_2022}. 

In this paper, we assess the feasibility of using generalized, ``multivalent" electronic structure for photon cycling and optical control of polyatomic molecules. For the purposes of this manuscript, we define ``multivalent" as describing systems with multiple valence electrons localized on the molecular OCC, in contrast to ``monovalent" systems with a single OCC-localized valence electron. We find that the bonding paradigms needed to engineer multivalent OCCs in polyatomic molecules are significantly different from the structural features previously used to design monovalent candidates. Our resulting approach is, to our knowledge, the first molecular design for OCCs that leverages orbital repulsion and \textit{geometric stabilization}, rather than solely bond ionicity and hybridization, to achieve quasi-closed cycling transitions. 

As proof-of-principle, we theoretically examine polyatomic molecules functionalized with $p$-block elements from group 13, 14, and 15 of the periodic table, such as Al. Experimental studies have already found diatomic analogs, namely AlF \cite{truppe_spectroscopic_2019,hofsass_optical_2021}, AlCl \cite{daniel_spectroscopy_2021}, and TlF \cite{hunter_prospects_2012,norrgard_hyperfine_2017}, to be excellent laser cooling candidates, and theoretical work  has identified around a dozen other promising species composed of $p$-block elements bonded to a halogen atom \cite{wells_electronic_2011,gao_laser_2015,wan_laser_2016,yang_ab_2016,gao_vibrational_2017,zhang_theoretical_2017,yuan_laser_2018,yuan_laser_2018,ren_configuration_2021,xia_theoretical_2017,zhang_ab_2018,li_theoretical_2020}. Functionalizing larger, polyatomic molecules with multivalent OCCs would combine previously heterogeneous features in a single molecule: 1) clock-state metrology and multi-electron degrees of freedom and 2) custom internal structure from the molecular ligand, which can yield long-lived, highly polarizable states \cite{kozyryev_precision_2017,yu_probing_2021}, tunable long-range interactions \cite{wall_simulating_2013, wall_realizing_2015,yu_scalable_2019}, and built-in co-magnetometers \cite{hutzler_polyatomic_2020,kozyryev_precision_2017}. However, the bonding paradigms which work to create cycling centers on monovalent AEL-type OCCs, such as substituting a halogen for a hydroxide \cite{kozyryev_proposal_2016, isaev_polyatomic_2016, ivanov_towards_2019, hutzler_polyatomic_2020, mitra_study_2021}, do not apply to these new systems.  For example, AlF has a structure which is highly amenable to photon cycling \cite{truppe_spectroscopic_2019,hofsass_optical_2021},  while AlOH does not (see Sec. \ref{sec:vibronic}).  

Thus, we must devise alternative approaches for identifying species which combine the advantages of polyatomic structure  with multivalent cycling centers. By choosing a linker atom which creates a more covalent metal-ligand bond than oxygen (such as sulfur) we find that molecular vibrations are decoupled from the valence OCC electrons through an intricate interplay of orbital hybridization, ionicity, and repulsion. We eluciate these orbital mechanisms for a variety of OCCs and ligands in order to gain insight into their effects on the internal structure and photon cycling in our candidate systems.  These results in turn enable us to deduce new bonding principles and optimal linker atom architectures for engineering optically controllable polyatomic molecules with complex electronic structure.

The model systems we characterize are of the form M$X$H, where M is a group 13, 14, or 15 atom and $X$ is a chalcogen ($X=$ O, S, Se, Te, Po) linker atom. Despite their increased structural complexity and challenging design constraints, our theoretical analysis predicts that several of these polyatomic molecules have highly decoupled, visible wavelength and near-infrared electronic transitions that support quasi-closed photon cycling schemes. For each class of polyatomics, we find species with diagonal Franck-Condon behavior, which in some cases, may enable photon cycling schemes that are quasi-closed up to one-part-in-10$^5$.

\begin{figure}
	\includegraphics[width=\columnwidth]{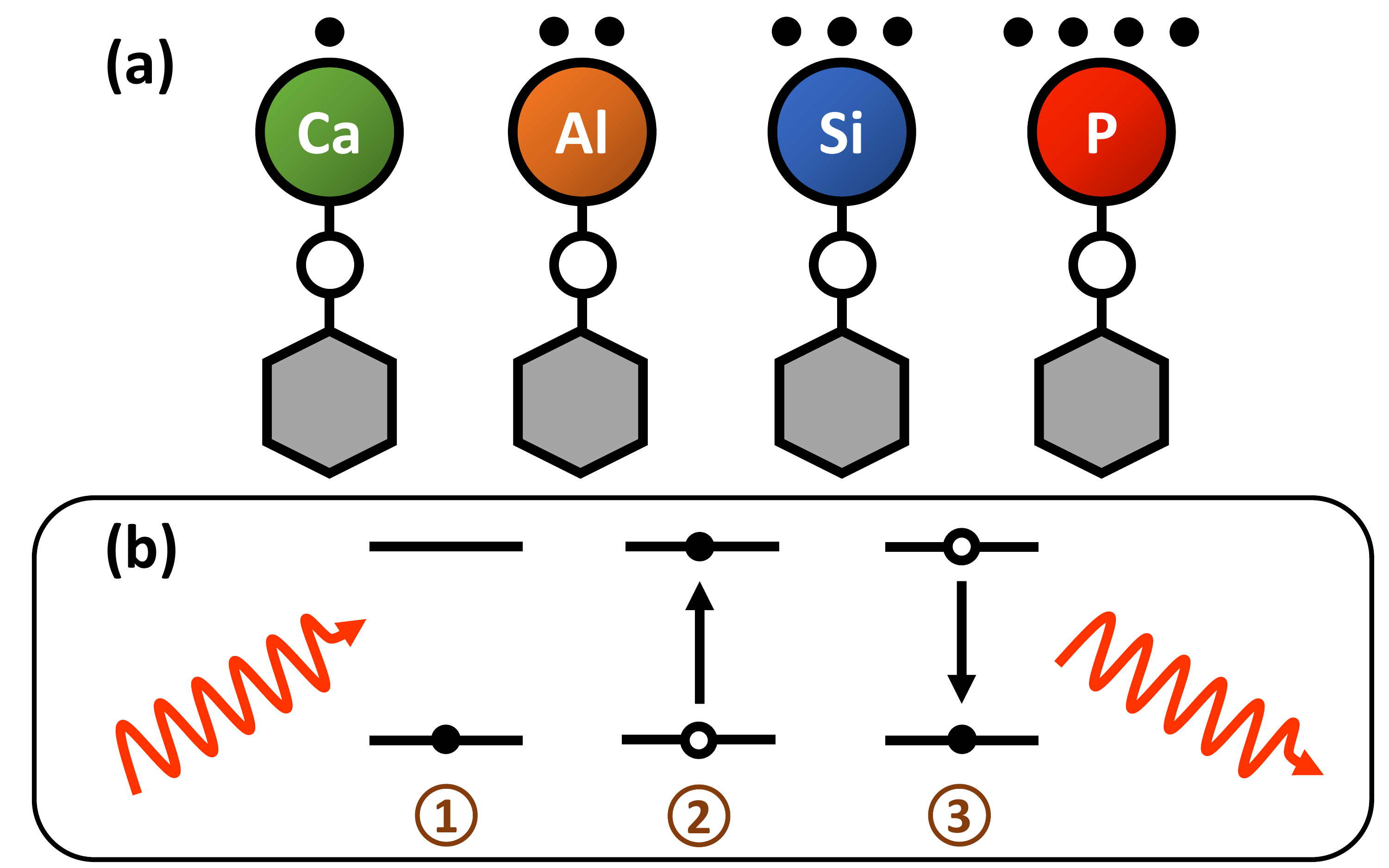}
	\caption{(a) Lewis dot structures depicting model ``monovalent" and ``multivalent" polyatomics with metal and metalloid optical cycling centers bonded to a pseudohalogen ligand. (b) Photon cycling in an idealized two-level system proceeds via the (1) absorption of a resonant photon, (2) followed by a transition to an excited electronic state. After the excited state lifetime elapses, (3) the excited state decays, releasing a photon through spontaneous emission.}
	\label{fig:valencediagram}
\end{figure}

The manuscript is organized as follows.  In section \ref{sec:pathways}, we review some of the general principles behind photon cycling, optical control, and laser cooling, while in section \ref{sec:methods}, we briefly describe our computational methods. In section \ref{sec:vibronic}, we examine the electronic and vibrational structure of several types of molecule, each of which has multiple cycling electrons and discuss the bonding principles which give rise to optical cycling. In section \ref{sec:design}, we further explore the bonding of these molecules by considering the effect of different metals and ligands on the structure of multivalent molecules.  Finally, in section \ref{sec:outlook}, we discuss the outlook for experimental study of these molecules and their potential applications.

\section{Pathways to Photon Cycling}\label{sec:pathways}
During photon cycling, valence electrons hosted by metallic OCCs undergo rapid cycles of coherent absorption and spontaneous emission of photons \cite{tarbutt_laser_2019, mccarron_laser_2018, fitch_laser-cooled_2021}, which can enable efficient optical state preparation, as well as high-fidelity detection and control.  The resulting momentum transfer, in analogy to atomic laser cooling \cite{chu_manipulation_1998}, can also facilitate slowing, cooling, and trapping of the entire molecule. Building molecules with properties amenable to cycling, however, is a challenging task. For instance, complex rovibrational structures in polyatomic molecules can serve as ``dark states" that interrupt an otherwise idealized two-level system for photon cycling.  Indeed, laser-coolable molecules follow a strict set of requirements on their internal structure \cite{di_rosa_laser-cooling_2004,isaev_polyatomic_2016,stuhl_magneto-optical_2008}, which include 1) intense visible or near-visible electronic transitions for photon cycling, 2) highly diagonal rovibrational decays and Franck-Condon factors, and 3) the absence of perturbing electronic states in the photon cycling pathway.

Establishing a highly closed photon cycling scheme requires detailed knowledge of transition energies and intensities between the cycling states and possible decay pathways to rovibrational dark states. Structural relaxation that accompanies spontaneous emission, in particular, will induce branching to vibrational substates, requiring additional re-pump lasers to restore population in the vibrationless cycling states. Most small, optically active molecules -- especially of low symmetry -- have vibronic wavefunctions that are separable under the Born-Oppenheimer approximation (i.e. the vibrational wavefunction can be expressed as independent of the electronic coordinates). Vibrational decays can therefore be predicted to high accuracy by computing Franck-Condon factors (FCF), which are defined as the overlap integral between vibrational wavefunctions $\psi_{v'}$, $\psi_{v''}$:
\begin{align}
    q_{v',v''}=\bigg|\int \psi_{v'}(\mathbf{Q'})\psi_{v''}(\mathbf{Q''})d\mathbf{Q}\bigg|^2\,.\label{eq:FCF}
\end{align}
Here, $v', v''$ are the vibrational quanta and $\mathbf{Q}$ are the nuclear coordinates of the normal modes. The vibrational branching ratios (VBRs) differ slightly from the FCFs due to the wavelength dependence of the spontaneous emission rate. They are defined as $\text{VBR}=\sum_{i, j} (\omega_{v_i', v_j''}^3\,q_{v_i',v_j''})/(\omega_{k}^3\,q_{k})$, where $\omega_{v', v''}$ is the transition wavelength between $|v''\ket\to |v'\ket$ and the FCFs $q$ are normalized over all decays. 

Typically, $10-10^2$ photons are needed for realizing high fidelity optical state preparation, readout, and state control. Similar numbers of photons can also be utilized for radiative deflection \cite{kozyryev_radiation_2016,kozyryev_coherent_2018}, steering, and confinement \cite{baum_1d_2020, augenbraun_laser-cooled_2020} of a cryogenic molecular beam. For laser slowing and capture of a small polyatomic molecule, up to $10^4-10^5$ photons are typically needed. This threshold, however, can be decreased significantly via indirect slowing and cooling methods, such as Stark/Zeeman deceleration \cite{hudson_efficient_2004,sawyer_magnetoelectrostatic_2007, hudson_deceleration_2009,petzold_zeeman_2018, aggarwal_deceleration_2021}, optoelectric slowing and cooling schemes \cite{zeppenfeld_sisyphus_2012,prehn_optoelectrical_2016}, as well as Zeeman-Sisyphus slowing  \cite{fitch_principles_2016, augenbraun_Zeeman-Sisyphus_2021}, which can precede direct loading into a magnetic trap \cite{lu_magnetic_2014}. Magnetically assisted approaches to slowing and trapping may be especially well-suited for multi-electron OCCs due to the presence of high-spin ground and metastable electronic states, as further discussed in Sec. \ref{sec:outlook}.

As we shall see, the $p$-block metals we consider make molecules which are bent.  Unlike the highly symmetric species that have been previously laser-cooled, the molecules we consider are at most $C_s$ symmetric and classified as asymmetric top molecules (ATM). ATMs, which possess three unequal moments of inertia ($I_A\neq I_B\neq I_C$), have electronic bands that can be categorized as $a$-type, $b$-type, or $c$-type, depending on the orientation of the transition dipole moment (TDM) relative to the molecule's three principal axes ($a$, $b$, $c$) (see Fig. \ref{fig:fig3} and Table \ref{tab:FCF}).  Each band has approximate angular momentum selection rules that can be leveraged to realize rotationally closed repump schemes with a manageable number of sidebands, as was shown in \cite{augenbraun_molecular_2020} for monovalent ATMs. This approach readily extends to multivalent ATMs, and a detailed discussion can be found in Appendix \ref{appendix:rotation}. 

Monovalent ATMs with optical cycling centers based on alkaline-earth metals have been previously considered \cite{augenbraun_molecular_2020}, and suitable ligands were found to maintain optical cycling characteristics despite their lower symmetry. In the systems we consider, we find that using an ATM structure is in fact generally \textit{necessary} for designing OCCs based on $p$-block metals, as linear analogs broadly appear to fail (see Appendix \ref{appendix:vibronic}).

\section{Computational Approach}\label{sec:methods}

We proceed by performing \textit{ab initio} analyses of the electronic and rovibrational structure of several polyatomics of the M$X$H form. The molecular candidates that we consider have three distinct typologies, with singlet, doublet, or triplet spin multiplicities in the ground state. This organization generally corresponds to a group 13 (divalent), 14 (trivalent), or 15 (quadrivalent) optical cycling center, respectively, attached to a pseudohalogen. 

The systems we consider possess ground states that at structural equilibrium are dominated by a single electronic configuration, making them ideally suited for analysis using coupled cluster methods \cite{lee_achieving_1995}. Calculations of the ground states are performed using coupled cluster with singles and doubles (CCSD), and excited states are characterized using analogous equation-of-motion schemes (EOM-CCSD). EOM-CC approaches, which are rigorously size-extensive, allow for multiconfigurational descriptions of target states within a single-reference formalism \cite{krylov_equation--motion_2008} and have been previously validated \cite{ivanov_towards_2019, li_emulating_2019, mengesha_branching_2020, zhang_accurate_2021, zhang_inner-shell_2022, lasner_vibronic_2022} for predicting the properties of a broad range of laser-coolable diatomic and polyatomic molecules. In this work, the traditional EOM excitation energies scheme (EOM-EE-CCSD) \cite{stanton_equation_1993} is used to study molecules with ground state singlet and triplet configurations (e.g. AlSH and PSH) from a singlet reference wavefunction, while spin-flip (EOM-SF-CCSD) \cite{krylov_size-consistent_2001,krylov_spin-flip_2006} is utilized for targeting states from a high-spin quartet reference (e.g. SiSH).

All electronic structure calculations are performed using the QChem 5.4 package \cite{epifanovsky_software_2021}, with wavefunction analyses conducted via the \textit{libwfa} library \cite{plasser_new_2014}.  Harmonic FCFs including Duschinsky rotation are computed using the \textit{ezFCF} code \cite{gozem_ezspectra_2021}.  Correlation-consistent sets of aug-cc-pVTZ(-PP) quality \cite{woon_gaussian_1993, wilson_gaussian_1999, peterson_systematically_2003} are used for calculations of single point energies, geometries, frequencies, and transition intensities.  For atoms heavier than period 3, core electrons are modeled using Stuttgart-type small core pseudopotentials (ECP10MDF, ECP28MDF, ECP60MDF) \cite{metz_small-core_2000, metz_small-core_2000-1}. Spin-orbit matrix elements are calculated perturbatively in the QChem code using the Breit-Pauli (BP) Hamiltonian \cite{pokhilko_spin-forbidden_2019, pokhilko_general_2019, epifanovsky_spin-orbit_2015}, for which we utilize relativistically contracted all-electron atomic natural orbital (ANO-R0) sets \cite{zobel_ano-r_2020} on the metal and metalloid cycling centers. Prior study \cite{cheng_perturbative_2018} has found that BP approaches -- despite excluding non-perturbative relativistic effects --  are able to capture dominant spin-orbit contributions, even in period 6 and 7 systems.

\section{Vibronic Structure}\label{sec:vibronic}

\subsection{Singlet Ground States: Group 13}
The simplest multivalent case we consider is a singlet system that arises from bonding a group 13 (i.e. B, Al, Ga, In, Tl) center to a pseudohalogen ligand.  Among diatomic molecules, AlF \cite{truppe_spectroscopic_2019,hofsass_optical_2021} and AlCl \cite{daniel_spectroscopy_2021} have strongly electronegative bonds and highly diagonal cycling transitions. One might expect, as with the alkaline earth series of polyatomics, that replacing the halogen atom with electronegative pseudohalogens, such as hydroxide (-OH), cyanide (-CN), ethynyl (-CCH), or boron dioxide (-OBO) ligands, would yield similarly laser coolable molecules. This turns out to be not the case (see Appendix \ref{appendix:vibronic}). Instead, we find that strongly bent molecules containing ligands with less electronegative character, such as hydrosulfide (-SH), do possess suitable bonding and diagonal cycling transitions.  Evidently, the unique orbital hybridization that enables laser cooling in AEL species does not universally translate to cycling centers from other columns of the periodic table.  

\begin{figure}
	\includegraphics[width=\columnwidth]{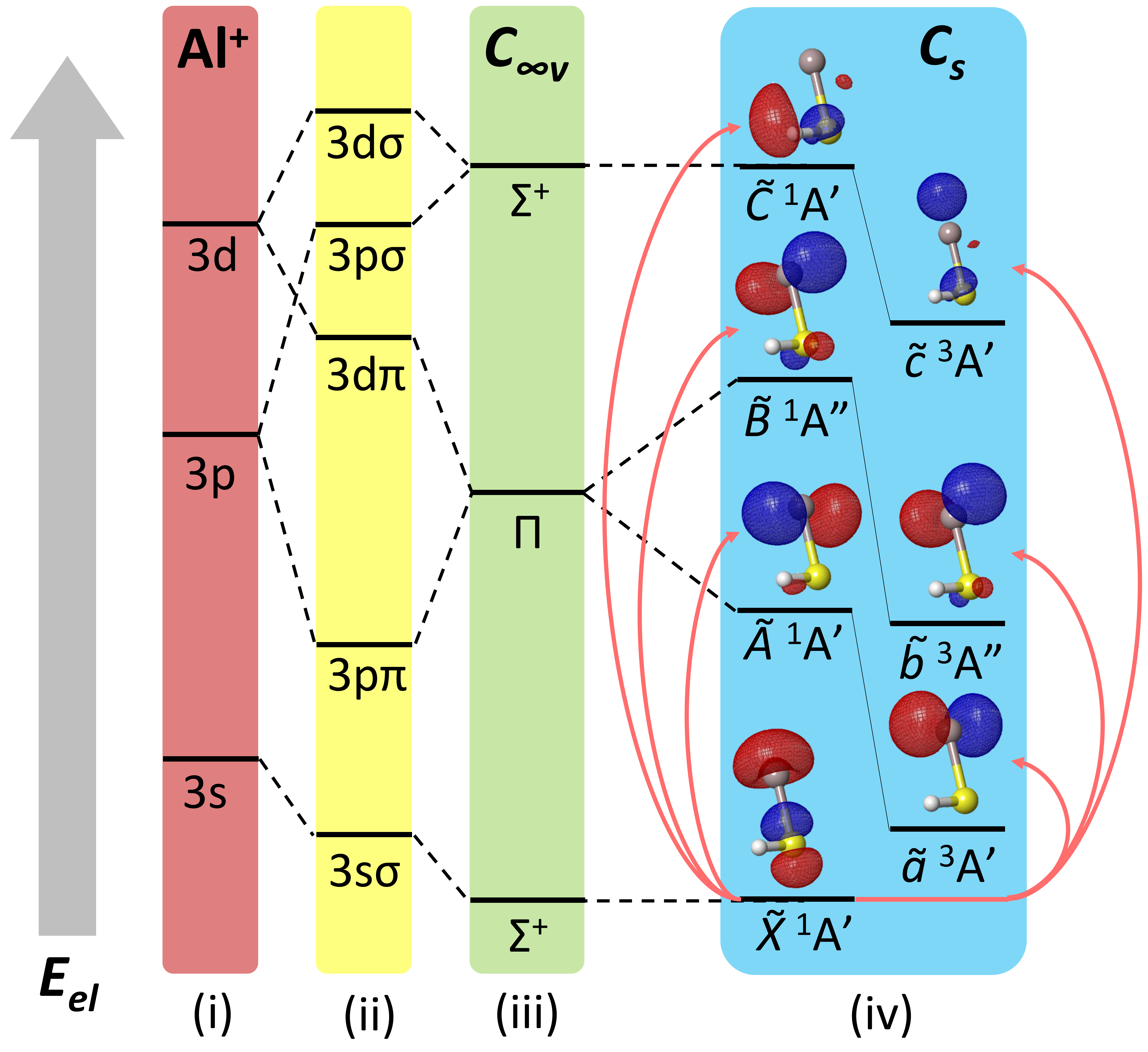}
	\caption{Ligand field diagram for low-lying electronic states of AlSH, not to scale. States are arranged bottom-up by increasing electronic energy ($E_{el}$). On the farthest left are (i) Al$^+$ cation shells, which are then split by the ligand (-SH$^-$) field into (ii) molecule-frame projections of orbital angular momentum ($\Lambda=\sigma, \pi, \delta, ...$). These orbitals mix to yield the electronic manifolds in the (iii) $C_{\infty v}$ linear and (iv) $C_s$ bent limits of the molecule.  Above each excited manifold in the bent case are natural transition orbitals (isovalue = 0.05) from the ground state computed using EOM-EE-CCSD. See Appendix \ref{appendix:vibronic} for a complete list of molecular orbital (MO) correlation diagrams and natural transition orbitals for relevant species.}
	\label{fig:nto_ALSH}
\end{figure}

\begin{figure}
    \includegraphics[width=\columnwidth]{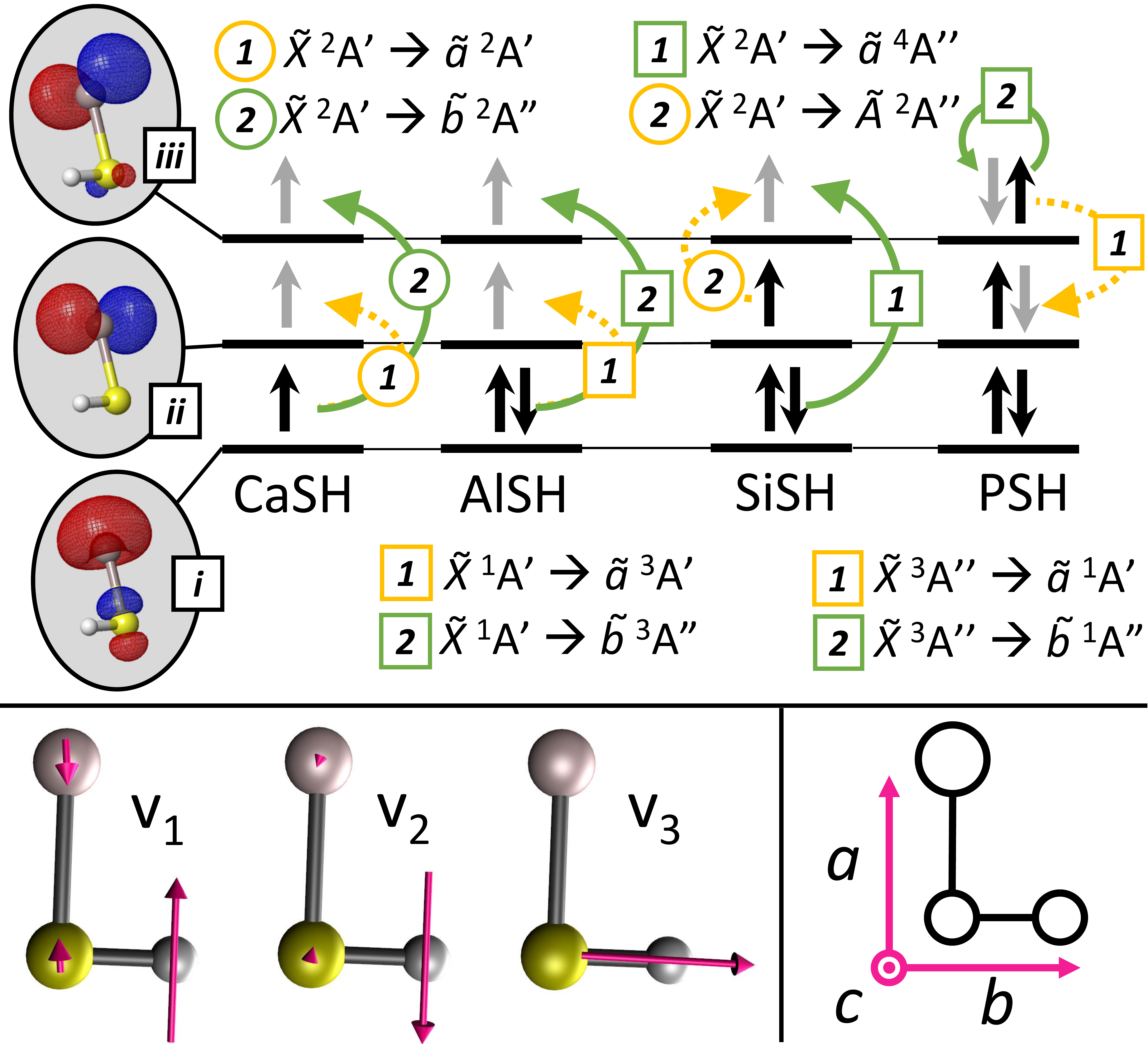}
	\caption{Top: Electronic configuration for ground and low-lying excited states of group 2 (CaSH~\cite{augenbraun_molecular_2020}), group 13 (AlSH), group 14 (SiSH), and group 15 (PSH) molecules. Depicted in the gray ovals on the left are frontier metal-centered MOs approximately corresponding to (i) M$^+$(s$\sigma$), (ii) M$^+$(p$\bar{\pi}$), and (iii) M$^+$(p$\pi$) atomic orbitals. Dashed yellow and solid green colored arrows depict (1) cycling and (2) intermediate decay channels, where circle and square labels indicate spin-allowed and spin-forbidden transitions. Black arrows in each MO denote electronic spins in ground state configuration, while gray arrows denote electronic spins upon excitation. See Fig. \ref{fig:nto} in Appendix \ref{appendix:vibronic} for complete MO and NTO schematics of low-lying electronic transitions.
	Bottom left: Vibrational modes roughly described as M-S stretch ($v_1$), M-S-H bend ($v_2$), and S-H stretch ($v_3$) in a multivalent asymmetric top molecule. Bottom right: Rotational axis convention for asymmetric top molecules (see Appendix \ref{appendix:rotation}). By convention, the tuple $(\hat{a},\hat{b},\hat{c})$ maps to the unit vectors $(\hat{z}, \hat{x}, \hat{y}$) used to label atomic orbitals.}\label{fig:fig3}
\end{figure}

A model system that we consider is the multivalent polyatomic alumininum monohydrosulfide (AlSH). In the ground state configuration\footnote{In this work, we adhere to spectroscopic conventions for labeling electronic states. $\tilde{X}$ is always the ground state; excited states with the same spin multiplicity as $\tilde{X}$ are $\tilde{A},\tilde{B},\ldots$, whereas those with different spin multiplicity are $\tilde{a},\tilde{b},\ldots$, both ordered in increasing energy.  The spin multiplicity is the superscript after the state name.  $A^p$ indicates the symmetry of the state, with $A'$ ($A''$) indicating that the electronic wavefunction is in-plane (out-of-plane) as shown in Fig. \ref{fig:nto_ALSH}.} ($\tilde{X}^1A'$), the Al-S bond is partially ionic, with a Mulliken charge ($Q_M$) of $+0.23$ on the metal and $-0.38$ on the sulfur. This corresponds to the withdrawal of a single $sp$-hybridized valence electron from the Al atom, leaving an Al($3s\sigma$) lone pair polarized away from the bond. The S-H bond is almost orthogonal ($\sim 90.19$\degree) to the Al-S bond, forming a prolate asymmetric top with $C_s$ symmetry.  As we shall see, this bond angle is a very important feature. The optimized geometry has three normal modes, which approximately correspond to Al-S stretch ($v_1$), Al-S-H bend ($v_2$), and S-H stretch ($v_3$), which are depicted in Fig. \ref{fig:fig3}.

The lowest three triplet states are $2.3$ to $4.2$ eV above the ground state and roughly correspond to the excitation of a single Al-localized $3s\sigma$ electron to $3p\pi+3d\pi$ and $3p\sigma+3d\sigma$ orbitals.  The $3s\sigma\to 3p\pi+3d\pi$ excitation is split by the off-axis SH ligand into an in-plane $\tilde{a}^3A'$ and out-of-plane $\tilde{b}^3A''$ state, while the $3s\sigma\to 3p\sigma+3d\sigma$ excitation corresponds to an on-axis $\tilde{c}^3A'$ state, as indicated by the ligand field diagram and natural transition orbital (NTO) analysis in Fig. \ref{fig:nto_ALSH} (see also Fig. \ref{fig:nto} in Appendix \ref{appendix:vibronic}). An analogous progression is obtained for the singlet states, which are all at ultraviolet energies. A summary of computed origins, rovibrational energies, and optimized geometries for both sets of states is listed in the Supplemental Material. 

Out of these six lowest lying electronic states, which are computed using EOM-EE-CCSD, we find that the $\tilde{b}^3 A''\to \tilde{X}^1A'$ transition ($\Delta E\sim 2.74$ eV) provides extremely diagonal vibrational branching, with an FCF of $q_{0,0}>0.997$ on the main vibrationless line. Dominant off-diagonal decays to $\tilde{X}$ are to the first ($q\sim 10^{-3}$) and second quanta ($q\sim 10^{-4}$) of the $v_1$ stretch mode. The $\tilde{b}^3 A''\to \tilde{X}^1A'$ vibrationless decay has one of the highest predicted FCFs among polyatomic systems that have been experimentally or theoretically characterized; however, as discussed later, losses from branching to intervening electronic states are non-negligible for this molecule. 

This finding is in line with prior theoretical and spectroscopic investigations of the iso-electronic AlF \cite{truppe_spectroscopic_2019} and AlCl molecules \cite{daniel_spectroscopy_2021}, which also found exceptionally high FCFs between the $X^1\Sigma^+$ state and $\{a^3\Pi$, $A^1\Pi\}$ manifolds. Benchmarks of our theoretical approach are in good agreement with observed geometries, energies, and lifetimes of AlF,  AlOH, and AlSH (see Table \ref{tab:benchmark} in the appendix).

\begin{figure}
	\includegraphics[width=\columnwidth]{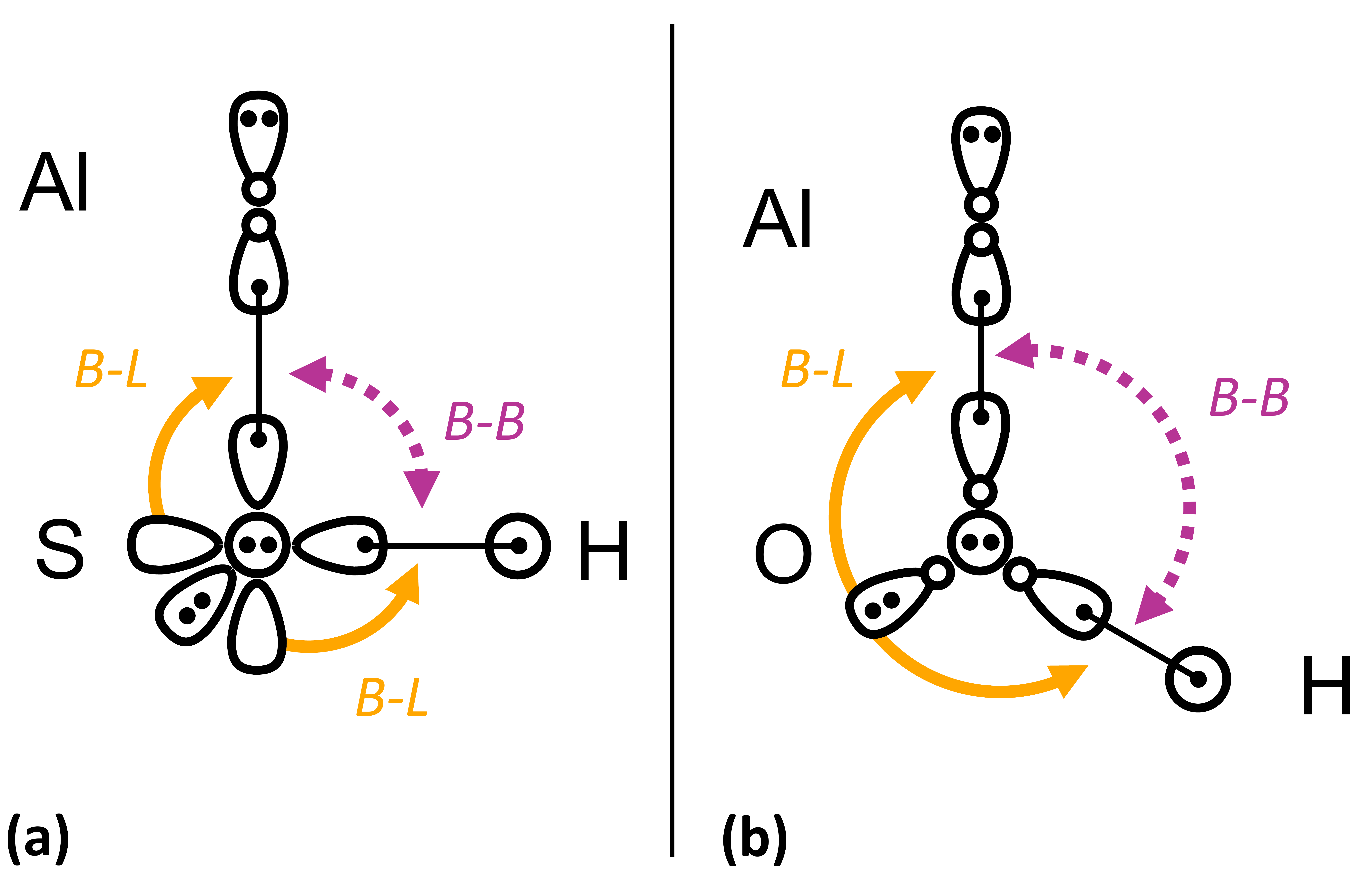}
	\caption{Simplified valence bonding (VB) diagram for (a) AlSH and (b) AlOH.  Arrows illustrate the two main repulsive effects involved in the geometry of the molecules: 1) the bond-bond repulsion between the aluminum-chalcogen and chalcogen-hydrogen bonds (purple dotted line, \textit{B-B}) and 2) repulsion from the in-plane $s$ lone pair on the chalcogen against the two bonds (yellow solid line, \textit{B-L}). In the absence of strong bond-bond repulsion, the bonds lock into the near-90\degree geometry provided by the orthogonal $p$-orbitals on the coordinating atom.  This the case in (b) AlSH, where the 3$p_z$ and 3$p_x$ valence lobe orbitals (with + and $-$ density components shown) on the coordinating sulfur bond to the H and to a sole 3$sp$-hybridized electron on the Al. The doubly occupied $sp$ orbital on the Al polarizes away from the Al-S bond, resulting in a 90\degree bond angle. Remaining out-of-plane $3p_y$ (circle with two dots) and in-plane $3s$ (teardrop with two dots) lone pairs are depicted on the sulfur atom. Low-lying excited states are formed by excited one of the two electrons in the doubly occupied Al(3$sp$) orbital to in-plane Al($3p_x$) and out-of-plane Al($3p_y$) orbitals. By contrast, in (b) AlOH, the short Al-O and O-H bonds lead to strong repulsion that opens the bond angle to $> $90\degree. This results in $sp$ hybridization of the valence orbitals on the O atom, while the $s$ lone pair builds in $p$-character. Additional details on this description can be found in Appendix \ref{appendix:vibronic}(2), which includes generalized valence bond (GVB) natural orbitals and GVB diagrams for all three molecular classes.}
	\label{fig:fig4}
\end{figure}

Meanwhile, the analogous $\tilde{b}^3A''\to \tilde{X}^1A'$ transition in AlOH has very non-diagonal Franck Condon factors, with $<30\%$ branching to the vibrationless ground state. There is an intuitive explanation for the diagonality of the $\tilde{b}^3 A''\to \tilde{X}^1A'$ transition in AlSH versus AlOH, based on qualitative arguments from valence bond (VB) theory \cite{goddard_description_1978} (see Fig. \ref{fig:fig4}). The larger size and lower electronegativity of S versus O results in increased bond lengths and reduced repulsion between the Al-S and S-H bonding electrons. The residual repulsion from the in-plane S($3s$) lone pair then dominates, causing AlSH to lock into the near-$90^\circ$ bent configuration given by the (orthogonal) bonding $p$ orbitals in S. We find that this geometry is stable when a valence electron around the Al atom is excited into an out-of-plane excited orbital, which is approximately decoupled from in-plane repulsive effects. 

By contrast, the shorter bond lengths in AlOH (induced by the electronegativity of the oxygen atom) cause increased bond-bond repulsion that pushes the bond angle past 90$^\circ$. This results in a bond angle that is highly sensitive to changes in orbital hybridization, and therefore highly non-diagonal Franck-Condon behavior. This VB picture is validated in Sec. \ref{sec:design} by high-level molecular orbital (MO)-based calculations, where we substitute even heavier atoms for S.

In contrast to the $\tilde{b}^3 A''\to \tilde{X}^1A'$ transition, the vibrationless transitions of AlSH from the other five excited manifolds to the ground state have either moderate ($< 0.7$) or poor ($<0.3$) FCFs (see Table \ref{tab:FCF}). The sub-optimal vibrational branching for the in-plane states $\{\tilde{a}^3A'$,  $\tilde{A}^1A'\}$ can be understood in terms of the repulsion between an in-plane Al$(3p\bar{\pi})$ lobe with the S-H bond, thereby opening the excited state bond angle to $\sim 100$\degree. This structural change drives vibrational branching to the bending mode ($v_2$) and Al-S stretch modes ($v_1$) during $\{\tilde{a}^3A', \tilde{A}^1A'\}\to \tilde{X}^1A'$ transitions. Conversely, UV excitations to the $\{\tilde{c}^3A', \tilde{C}^1A'\}$ states preserve the bond angle, but also significantly reduce the ionicity of the Al-S bond. This is evidenced by the reduced Mulliken charges on Al, which is 0.03 for $\tilde{c}^3A'$ and 0.08 for $\tilde{C}^1A'$. Consequently, these states exhibit substantially longer ($>28\%$) bond length and increased branching to the ground state $v_1$ mode. 

Globally, we also observe that the triplet states have shorter bond lengths and more diagonal FCFs than the excited singlet states. The $\tilde{B}^1A''\to\tilde{X}^1A'$ vibrationless decay, for instance, has an FCF of only $\sim 0.2$, with the primary $v_1$ loss attributable to an increased ($\sim 6\%$) Al-S bond length in the excited state. This effect can be rationalized as a consequence of spin-exchange effects between the frontier orbitals. In the low-lying $A'$ and $A''$ excited states, a single valence electron from the doubly occupied metal $s\sigma$ antibonding orbital is promoted to the $p\bar{\pi}$ and $p\pi$ antibonding orbitals, respectively. As the singly occupied $\sigma$ orbital is orthogonal to the singly occupied $p\bar{\pi}$ and $p\pi$ orbitals, spin-exchange interactions between the unpaired electrons in the triplet spin configurations contribute negatively \cite{lowdin_exchange_1962} to the many-electron energy and stabilize the molecular potential. By contrast, in the singlet spin configurations, spin-exchange interactions between the singly occupied, orthogonal orbitals contribute positively to the many-electron energy. Minimizing the molecular potential in singlet excited configurations therefore causes delocalization of the frontier $\pi$ orbitals and lengthening of the M-S bond.

\begin{figure}
	\includegraphics[width=\columnwidth]{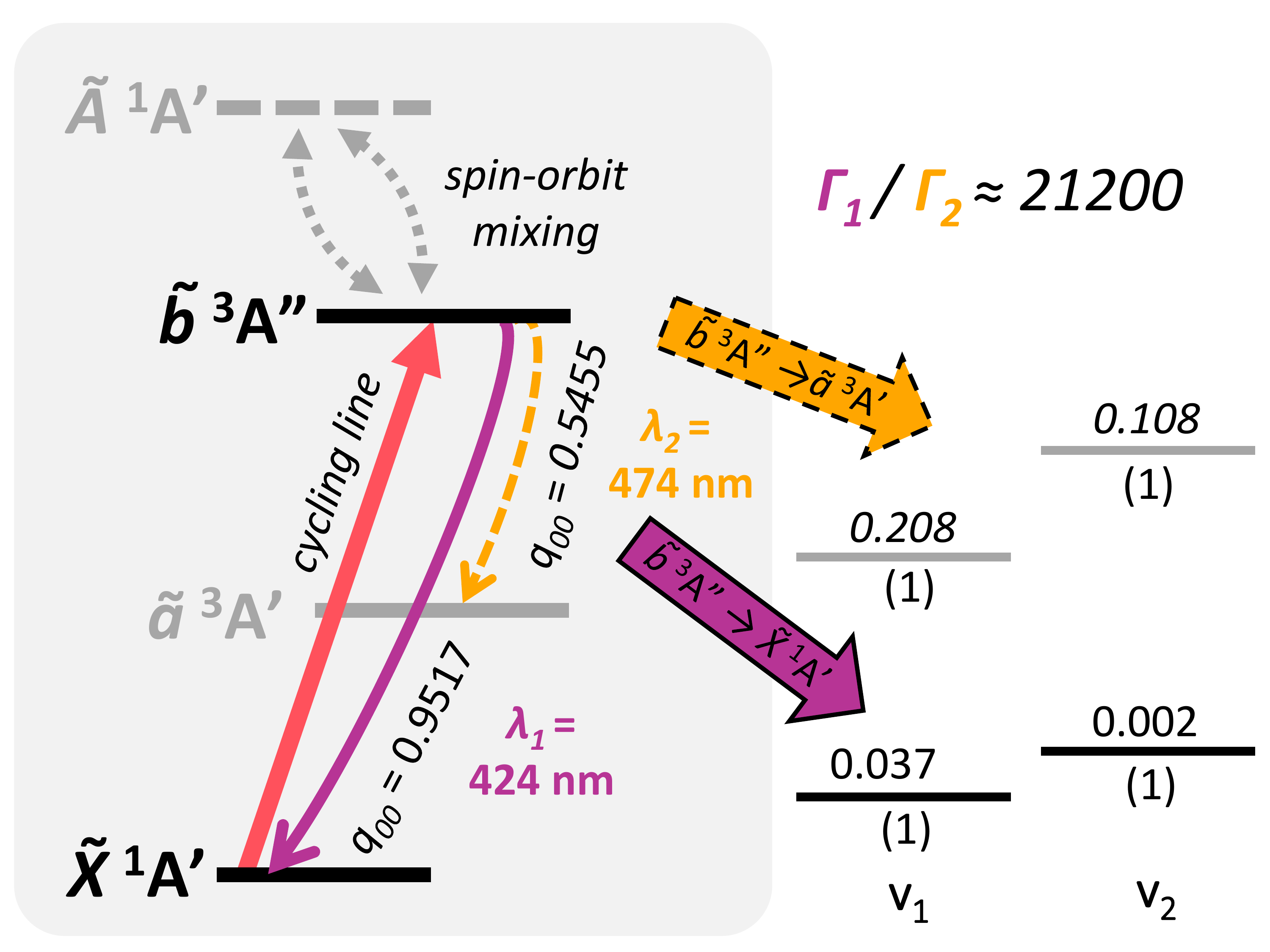}
	\caption{Generic photon cycling scheme for group 13 molecules, using vibronic branching data from InSH. The main cycling transition (red solid arrow) is from $\tilde X^1A'$ to $\tilde{b}^3A''$. On the left, solid curved (purple) and dashed curved (yellow) lines denote vibration-free decays to vibrational channels in the $\tilde X^1A'$ and $\tilde a^3A'$ manifolds, respectively, after spontaneous emission from the $\tilde{b}^3A''$ upper cycling state. Pairs of gray dashed arrows depict the spin-orbit induced mixing between the first excited singlet state $\tilde{A}^1A'$ and the upper $\tilde{b}^3A''$ cycling state. Levels on the right hand side depict leading off-diagonal FCFs for decays to the $\tilde a^3A'$ and $\tilde X^1A'$ manifolds, with corresponding transition wavelengths denoted $\lambda_2$ and $\lambda_1$, respectively. Decimals above the levels denote the Franck-Condon factors (eq. \ref{eq:FCF}) normalized relative to the respective electronic transition, while numbers underneath indicate the vibrational quanta in each mode ($v_i$). Due to spin-orbit effects from the In center, more than 99.995$\%$ of decays out the $\tilde {b}^3{A''}$ state connect directly to the $\tilde{X}^1A'$ state, as indicated by the suppression factor ($\Gamma_1/\Gamma_2$) in the upper right hand corner (see Table \ref{tab:FCF}). Analogous optical cycling schemes can be utilized for all group 13 species, and cycling schemes for other multivalent classes can be found in Fig. \ref{fig:BiSH_cycling} and \ref{fig:SiSH_cycling} in the Appendix. Note that level spacings are not drawn to scale.}
	\label{fig:lc_scheme} 
\end{figure}

The relevant vibronic level schematic for group 13 molecules, including AlSH is shown in Fig. \ref{fig:lc_scheme}. In the case of AlSH, the cycling scheme is centered around a $\sim$ 450 nm transition from $\tilde{X}^1A'$ to $\tilde{b}^3A''$. The upper $\tilde{b}^3A''$ state has $\sim 0.1\%$ and $\sim 0.02\%$ admixtures with the $\tilde{X}^1A'$ and $\tilde{A}^1A'$ manifolds due to spin-orbit coupling, resulting in an extremely narrow $2\pi\times 4.28$ Hz ($\mu\sim 2.81 \times 10^{-3}$ D) dipole-allowed transition from the ground state.\footnote{This can be compared against transition widths for AEL-type monovalent OCCs, which are typically $\sim 2\pi\times 1-10$ MHz.} Electronic decays from $\tilde{b}^3A''$ to the intermediate $\tilde{a}^3A'$ states are calculated to have a TDM of $\mu\sim2.58\times 10^{-2}$ D. Long transition wavelengths, however, suppress the intensity of this band by a ratio of at least 3.7:1 relative to the  $\tilde{b}^3A''\to\tilde{X}^1A'$ cycling transition. 

The long lifetime of the excited state would result in very low scattering rates, making Dopper laser cooling not feasible for AlSH.  However, since the transition dipole moment depends on the spin-orbit (SO) coupling, we can improve both the scattering rate and the $\tilde{b}^3A''\to\tilde{a}^3A'$ branching by choosing heavier group 13 cycling centers with increased SO effects. This naturally also leads to an increased suppression factor for decays to the intervening $\tilde{a}^{3}A'$ electronic state, as the intensity of the dipole-allowed intermediate $\tilde{b}^3A''\to \tilde{a}^3A'$ decays do not increase, and in fact, slightly decrease with the change to heavier metal centers. We find that substituting the Al cycling center with Ga and In atoms marginally decreases the vibrationless cycling line ($\tilde{b}^3A''\to\tilde{X}^1A'$) FCF to 0.9804 and 0.9517, respectively, while significantly increasing the SO-induced linewidths to $2\pi\times 137$ Hz ($\mu\sim 1.32\times10^{-2}$ D) and $2\pi\times 2.91$ kHz ($\mu\sim 6.67\times10^{-2}$ D). The suppression factor into $\tilde{a}^3A'$ similarly increases to $86.5:1$ GaSH and $21200:1$ for InSH. Scattering of $>10^{4}$ photons is therefore plausible before needing to repump out of the intermediate $\tilde{a}^3A'$ state for these heavier, isoelectronic species (see Supplemental Material for details).  Note, however, that branching at the $\lesssim 10^{-4}$ level can be induced by vibronic effects \cite{zhang_accurate_2021, paul_laser-induced_2019}, which are not considered here but warrant further study.

\begin{table*}
\caption{\label{tab:FCF}
Transition energies, wavelengths, vibrationless FCFs ($q_{00}$), linewidths, and band orientation (see Appendix \ref{appendix:rotation}) for cycling and decay transitions of MSH molecules. Data for intersystem lines assume intensity borrowing due to SO mixing.
}
(a) Singlet molecules
\begin{ruledtabular}
\begin{tabular}{llllllllllll}
& \multicolumn{5}{c}{Cycling Transition ($\tilde{b}^3A''\to\tilde{X}^1A'$)} & \multicolumn{6}{c}{Intermediate Decay ($\tilde{b}^3A''\to\tilde{a}^3A'$)} \\
\cline{2-6} \cline{7-12}
Species & $\Delta E$ (eV) & $\lambda$ (nm) & $q_{00}$ & $\Gamma$ (Hz) & Band & $\Delta E$ (eV) & $\lambda$ (nm) & $q_{00}$ & $\Gamma$ (Hz) & Band & Suppression \\
\hline
BSH & 2.396 & 517 & 0.9600 & $2\pi\times4.24$ & $ab$-type & 0.9350 & 1326 & 0.1708 & $2\pi\times 97.6$ & $c$-type & 0.0434\\

AlSH & 2.744 & 451 & 0.9974 & $2\pi\times4.28$ & $ab$-type & 0.4045 & 3065 & 0.5645 & $2\pi\times1.16$ & $c$-type & 3.70\\

GaSH & 3.113 & 398 & 0.9804 & $2\pi\times 137$ & $ab$-type & 0.4143 & 459 & 0.3658 & $2\pi\times 1.58$ & $c$-type & 86.5\\

InSH & 2.921 & 424 & 0.9517 & $2\pi\times 2910$ & $ab$-type & 0.306 & 474 & 0.5455 & $2\pi\times 0.137$ & $c$-type & 21200\\

TlSH & 3.330 & 372 & 0.5210 & $2\pi\times 681000$ & $ab$-type & 0.2297 & 400 & 0.2706 & $2\pi\times 0.00996$ & $c$-type & 6.83$\times10^7$\\

\end{tabular}
\end{ruledtabular}
\vspace{0.25 cm}

(b) Doublet molecules
\begin{ruledtabular}
\begin{tabular}{llllllllllll}
& \multicolumn{5}{c}{Cycling Transition ($\tilde{a}^4A''\to\tilde{X}^2A'$)} & \multicolumn{6}{c}{Intermediate Decay ($\tilde{a}^4A''\to\tilde{A}^2A''$)} \\
\cline{2-6} \cline{7-12}
Species & $\Delta E$ (eV) & $\lambda$ (nm) & $q_{00}$ & $\Gamma$ (Hz) & Band & $\Delta E$ (eV) & $\lambda$ (nm) & $q_{00}$ & $\Gamma$ (Hz) & Band & Suppression \\
\hline
CSH & 2.121 & 585 & 0.6667 & $2\pi\times 37.8$ & $ab$-type & 1.036 & 1143 & 0.3412 & $2\pi\times0.0298$ & $c$-type & 1268\\

SiSH & 2.654 & 467 & 0.7049 & $2\pi\times 192$ & $ab$-type & 2.090 & 593 & $0.7515$ & $2\pi\times 0.219$ & $c$-type & 875\\

GeSH & 2.845 & 436 & 0.2498 & $2\pi\times 770$ & $ab$-type & 2.319 & 535 & $0.5631$ & $2\pi\times 2.37$ & $c$-type & 325

\end{tabular}
\end{ruledtabular}

\vspace{0.25 cm}

(c) Triplet molecules
\begin{ruledtabular}
\begin{tabular}{llllllllllll}
& \multicolumn{5}{c}{Cycling Transition ($\tilde{b}^1A''\to\tilde{X}^3A''$)} & \multicolumn{6}{c}{Intermediate Decay ($\tilde{b}^1A''\to\tilde{a}^1A'$)} \\
\cline{2-6} \cline{7-12}
Species & $\Delta E$ (eV) & $\lambda$ (nm) & $q_{00}$ & $\Gamma$ (Hz) & Band & $\Delta E$ (eV) & $\lambda$ (nm) & $q_{00}$ & $\Gamma$ (Hz) & Band & Suppression \\
\hline
PSH & 0.8558 & 1449 & 0.9018 &  $2\pi\times 8.86 \times 10^{-4}$ & $c$-type & 0.5307 & 2336 & 0.2986 & $2\pi\times 14.4$ & $c$-type & $6.13\times 10^{-5}$\\

AsSH & 0.8524 & 1455 & 0.9224 & $2\pi\times 0.900$ & $c$-type & 0.3773 & 3286 & 0.2690 & $2\pi\times 3.53$ & $c$-type & 0.255\\

SbSH & 0.7897 & 1570 & 0.9572 & $2\pi\times 2.46$ & $c$-type & 0.1969 & 6297 & 0.5130 & $2\pi\times 0.214$ & $c$-type & 11.5\\

BiSH & 0.7750 & 1600 & 0.9674 & $2\pi\times 6.19$ & $c$-type & 0.117 & 10566 & 0.3994 & $2\pi\times0.0308$ & $c$-type & 200\\

\end{tabular}
\end{ruledtabular}
\end{table*}

\subsection{Doublet Ground States: Group 14}
Next, we examine neutral polyatomic systems with group 14 (e.g. C, Si, Ge) optical cycling centers bonded to a hydrosulfide ligand (-SH). The molecules in this class have doublet ground states similar to monovalent alkaline earth-pseudohalogen systems, but a much larger valence space that includes ground state electron occupation in $p$-orbitals as well as $s$-orbitals localized to the optical cycling centers. 

A model system is SiSH. In its ground state, the Si atom has two unpaired electrons in $3p_z$ and $3p_x$ orbitals, yielding a $(3s)^2(3p_z)(3p_x)$ valence configuration. Like AlSH, the ground state of SiSH consists of a bond between an unpaired Si$(3p\sigma)$ orbital, which has Si$(3p_z)$ character, and the unpaired SH $\sigma$-electron. The remaining in-plane $3p\bar{\pi}$ orbital Si$(3p_x)$ contains one unpaired electron and is the frontier orbital for a $^2A'$ electronic manifold. As before, the residual Si$(3s)$ lone pair electrons polarize against the bond by mixing in negative $3p\sigma$ character. Quartet configurations can be obtained by exciting one of the Si$(3s)$ electrons into the out-of-plane Si$(3p\pi)$ orbital, which has Si$(3p_y)$ character, to obtain a $^4A''$ state (see Fig. \ref{fig:fig3} and Fig. \ref{fig:nto} in Appendix \ref{appendix:vibronic}). This configuration is analogous to high-spin states ($^4\Sigma^-$) that have been spectroscopically observed in diatomics such as CF \cite{grieman_a4x_1983}, SiF \cite{verma_4sigma--2pi_1962, martin_4sigma--x2pi_1973}, and GeF \cite{martin_rotational_1973}. Calculations for group 14 molecules are performed via EOM-SF-CCSD, using this high-spin quartet $^4A''(m_s=3/2)$ reference to target low-spin $m_s=0$ states.

The ground state molecular geometry of SiSH is bent, with a bond angle of $\sim 100$\degree. In analogy to earlier arguments, the larger bond angle can be attributed to repulsion between the S-H bond and the in-plane Si$(p\bar{\pi})$ orbital. The Si-S bond is polar, with a Mulliken's charge of $Q_M=+0.091$ on the cycling center and $Q_M=-0.279$ on S. Immediately above the ground state is a low-lying out-of-plane $\tilde{A}^2A''$ state (0.56 eV) and in-plane $\tilde{B}^2A'$ state (3.79 eV), which corresponds to excitations from the in-plane Si($3p\bar{\pi}$) to the out-of-plane Si($3p\pi$) and Si($3p\sigma$) orbitals, respectively. A high-spin $\tilde{a}^4A''$ state (2.65 eV) with occupation in an out-of-plane Si($3p\pi$) orbital is predicted between the two doublet excited states. Above all three states is the $\tilde{C}^2A'$ manifold (3.95 eV), which has occupation in Si($3d\sigma$).

The optimal cycling transition in this system is from the $\tilde{X}^2A'$ to the $\tilde{a}^4A''$ state. Between the $\tilde{a}^4A''$ and $\tilde{X}^2A'$ state is the low-lying intermediate $\tilde{A}^2A''$ state, which has non-diagonal decays to the ground state from the $\tilde{a}^4A''$ state. For the $\tilde{a}^4A''\to\tilde{X}^2A'$ cycling transition, the vibrationless FCF is $q_{0,0}\sim 0.7049$, while the leading off-diagonal decay to the first quantum of the Si-S stretch mode ($v_1$) has an FCF of $q\sim 0.2081$. Subleading off-diagonal losses at the percent-level include decays to the first quanta of the $v_2$ bending mode ($q\sim 5.04\times10^{-2}$) and second quanta of the $v_1$ Si-S stretch mode ($q\sim 1.92\times10^{-2}$). Despite a lower vibrationless FCF than the group 13 and 15 systems considered earlier, the sum of the leading two FCFs for the SiSH $\tilde{a}^4A''\to\tilde{X}^2A'$ transition exceeds $90\%$ and the sum of the leading four FCFs exceeds $98\%$, which is comparable to the leading FCFs of the most diagonal polyatomic systems.

Intersystem transitions to the $\tilde{a}^4A''$ state are allowed via a combination of direct spin-orbit mixing with the $X^2A'$ state and intensity borrowing from spin-allowed transitions to higher doublet states, resulting in kHz-scale scattering rates. Branching to the intermediate $\tilde{A}^2A''$ state is suppressed due to disfavored spin-orbit couplings by a factor of $\sim 900$ (see Table \ref{tab:FCF}). Substituting the Si center for a Ge atom increases the scattering rate and suppression factor by a factor of $\sim 3$ at the cost of a lower vibrationless FCF. We further find that substituent cycling centers heavier than Ge (i.e. Sn and Pb) do not provide stable geometries for photon cycling between the $\tilde{a}^4A''$ and $X^2A'$ states.

\subsection{Triplet Ground States: Group 15}
In this section, we consider group 15 (P, As, Sb, Bi) centers bonded to a hydrosulfide ligand (-SH) and find that this approach works well, despite possessing significantly different electronic configurations from the original group 13 prototype. Molecules with group 15 centers have ground triplet configurations, which provide for a unique set of properties to combine with optical cycling and polyatomic structure, including large magnetic moments and hyperfine states with widely tunable spin couplings in the ground electronic state.

We proceed with an analysis of the electronic structure of these systems, which is conducted using EOM-EE-CCSD with a singlet reference. A prototypical case is PSH. Unlike the Al atom, which only has a single unpaired $3p\sigma$ electron in its ground state, the P atom has three singly occupied $p$-orbitals, corresponding to a $(3s)^2(3p_x)(3p_y)(3p_z)$ high-spin valence configuration. The unpaired P($3p\sigma$) orbital, which has P($3p_z$) character,  pairs to an unpaired $\sigma$-orbital on the SH ligand, forming a polar covalent bond between the P ($Q_M=+0.052$) and S ($Q_M=-0.246$) atoms. The residual unpaired P($3p\bar{\pi}$) and P($3p\pi$) electrons, which have P($p_x$) and P($p_y$) character, form a $^3A''$ state. Meanwhile, the P($3s$) lone pair mixes in some negative P($3p\sigma$) character to polarize against the new P-S bond. Singlet configurations of PSH correspond to the singlet pairing of the singly occupied P($3p\bar{\pi}$) and P($3p\pi$) orbitals $(^1A'')$ plus the two states where either P($3p\bar{\pi}$) or P($3p\pi$) are doubly occupied (see Fig. \ref{fig:fig3} and Fig. \ref{fig:nto} in Appendix \ref{appendix:vibronic}). 

Due to nodal planes in the residual non-bonding P(3$p\pi$) and P(3$p\bar{\pi}$) orbitals, spin-exchange interactions favor a triplet over a singlet configuration in the ground state. The ground state of PSH therefore has the term $\tilde{X}^3A''$. In this state, PSH has a bond angle of $96^\circ$, slightly larger than that of AlSH. The lowest singlet states correspond to excitations from the out-of-plane P($3p\pi$) orbital to the in-plane P($3p\bar{\pi}$) orbital ($\tilde{a}^1A'$), the out-of-plane P($3p\pi$) ($\tilde{b}^1A''$), and the on-axis P($3d\sigma$) orbital ($\tilde{c}^1A'$). These singlet states have origins at 0.325 eV, 0.856 eV, and 3.962 eV, respectively. Triplet progressions to the in-plane and out-of-plane P($p\pi$) orbitals correspond to the $\tilde{A}^3A'$ (3.468 eV) and $\tilde{B}^3A''$ (3.740 eV) states.

Similar to our findings in the last section, we find that the $\tilde{X}^3A''\to\tilde{b}^1A''$ transition provides the most diagonal vibrationless FCF and is therefore well-suited as a cycling transition. The main vibrationless decay has an FCF of $q_{0,0}\sim 0.9018$, with dominant decays to the first ($q\sim 0.0907$) and second ($q\sim 2.506 \times10^{-3}$) quanta of the P-S stretch mode ($v_1$), the first ($q\sim 3.852\times10^{-3}$) quanta of the bending mode ($v_2$), and a $q\sim 6.621 \times 10^{-4}$ decay to a stretch-bend $(v_1=1, v_2=1)$ combination state. Decays from and to in-plane states (such as the intermediate $\tilde{a}^1A'$ state) yield less diagonal FCFs, due to analogous bonding principles discussed in the previous section. 

For group 15 systems, the spin-orbit interaction on the cycling center preferentially couples $\tilde{b}^1A''$ to in-plane triplet states over out-of-plane triplet states as a consequence of spatial selection rules (see Sec.\ref{sec:design}(B)). As the ground state is out-of-plane for this class of molecules, the intensity borrowing for the $\tilde{X}^3A''\to\tilde{b}^1A''$ cycling transition is weaker than in group 13 systems, resulting in significantly narrower cycling transitions (see Table \ref{tab:FCF}(b) and Sec. \ref{sec:design}(A)). As with group 13 molecules though, substitution of heavier cycling centers leads to broader cycling transitions and more heavily suppressed decays to the intermediate $\tilde{a}^1A'$ state. We also find that the vibrationless FCF on the cycling line is highly diagonal ($q_{0,0}> 90\%$) for all group 15 species and improves with heavier OCC substituents.


\section{Design principles}\label{sec:design}
\subsection{Linker atom and bond polarity}
The design of molecular OCCs requires cycling degrees of freedom to be decoupled from rovibrational modes of the molecule; that is -- the geometry of the molecule should not change upon excitation in the optical cycling scheme. Linker atoms have a significant influence on cycling characteristics in polyatomic systems, both by controlling the nature of the OCC-ligand bond, as well as spatially decoupling the rovibrational modes of the ligand from metal-centered cycling. Among conventional AEL-pseudohalogen systems, the MO\textit{R} motif, which utilizes an oxygen atom to link cycling centers ($M$) to a functional group ($R$), has seen enormous success in identifying and engineering laser-coolable systems \cite{isaev_polyatomic_2016,kozyryev_proposal_2016, isaev_laser-coolable_2017,kozyryev_sisyphus_2017,ivanov_towards_2019,li_emulating_2019,ivanov_toward_2020,kozyryev_determination_2019,klos_prospects_2020, baum_1d_2020,augenbraun_laser-cooled_2020, mitra_direct_2020,vilas_magneto-optical_2021,augenbraun_observation_2021, paul_electronic_2021,dickerson_franck-condon_2021,mitra_pathway_2022,zhu_functionalizing_2022}. Other linker paradigms (e.g. S, N, C) have also been explored for monovalent molecules \cite{isaev_polyatomic_2016,orourke_hypermetallic_2019,ivanov_two_2020}, particularly systems of reduced symmetry \cite{augenbraun_molecular_2020}.

In our analysis, we have considered the effects of a variety of chalcogens (X=O, S, Se, Te, Po) as coordinating atoms for optically active polyatomics. The lightest of all the possible choices is oxygen, which -- as mentioned earlier -- is widely used as a linker atom in monovalent, laser coolable polyatomics. Bonds that are coordinated to the $p$-electrons of oxygen naively adopt a perfect 90$^\circ$ angle, due to the orthogonality of the atomic orbitals. This geometry, however, is further altered by electrostatic mechanisms. In the case of oxygen, the electronegative character of the atom produces a highly polar metal-ligand bond, which has two effects on the structure of these molecules. First, the ionicity of the bond causes coulombic and bond-bond repulsion effects (which are optimal in the linear case) \cite{greetham_ultraviolet_2000, essers_scf_1982, ni_laser_1986} to overcome repulsion from the in-plane oxygen lone pair (which prefers an acute structure) \cite{barclay_millimeter-wave_1992}, resulting in a linear or highly symmetric nonlinear molecular geometry. Second, the highly polar metal-ligand bond polarizes the unpaired cycling electron away from the bond, decoupling it from the rest of the molecule. Indeed, it has been widely suggested in the cold molecule community that the existence of a highly polar bond between the OCC and ligand may be an important condition for diagonal FCFs and laser coolability  \cite{isaev_polyatomic_2016, kozyryev_proposal_2016, ivanov_towards_2019, li_emulating_2019, hutzler_polyatomic_2020, ivanov_search_2020, dickerson_franck-condon_2021,mitra_pathway_2022,zhu_functionalizing_2022}.

\begin{figure}
	\includegraphics[width=\columnwidth]{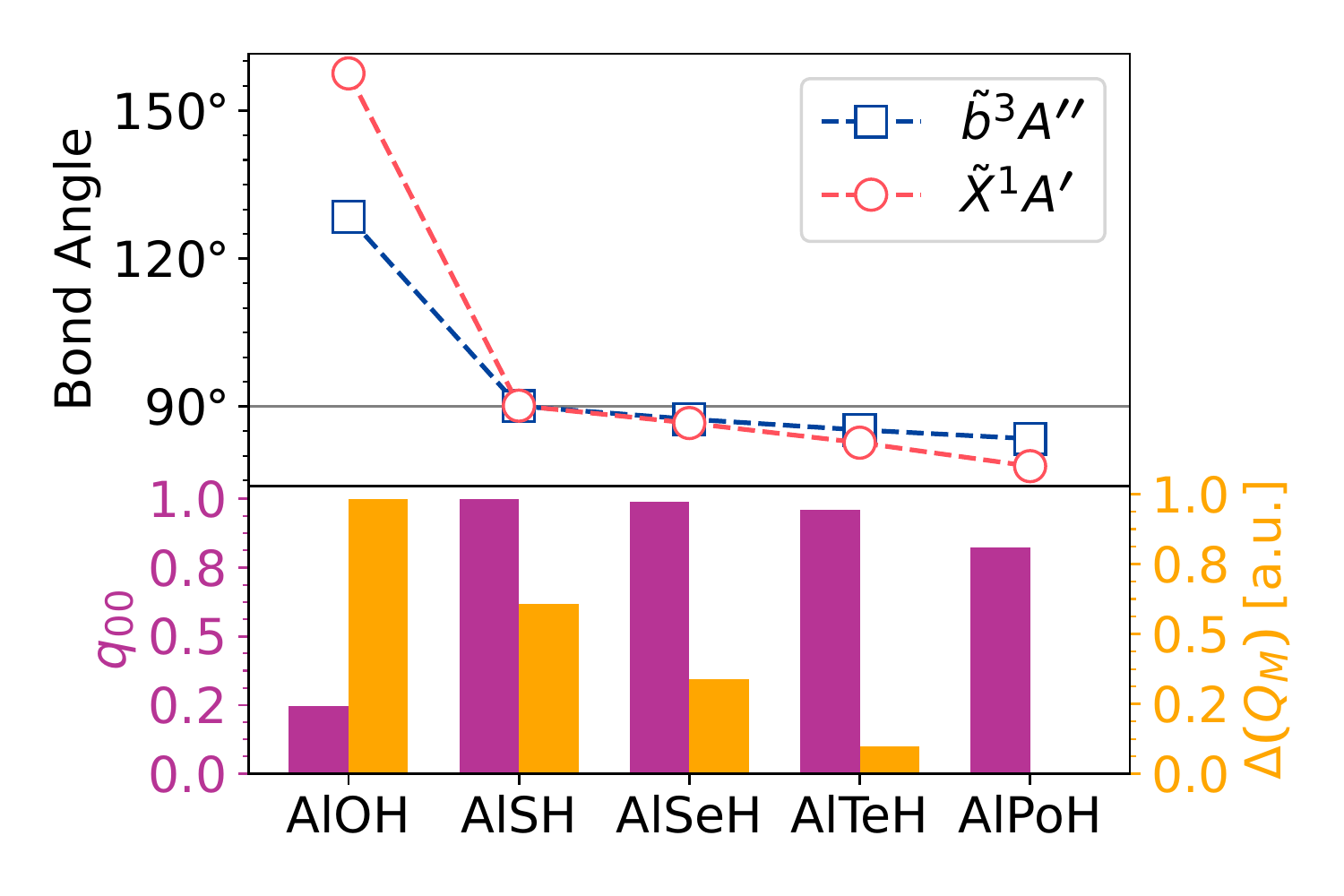}
	\caption{Effects of chalcogen substituents on Franck-Condon factors, molecular geometries, and bond polarity.  Line plot depicts bond angles for the cycling states $\tilde{X}^1A'$ and $\tilde{b}^3A''$ of five different Al$X$H molecular species with different chalcogen linker atoms [$X$=O, S, Se, Te, and Po]. Pairs of bar plots in the lower half of the panel indicate FCFs (left, purple) and Mulliken charge difference ($Q_M$, right, yellow) between the cycling center and linker atoms. A thin horizontal gray line on the upper plot denotes 90$^\circ$ bond angle.}
	\label{fig:periodic}
\end{figure}

By contrast, multivalent species with $p$-block OCCs and an oxygen linker (i.e. AlOH) have nonlinear geometries. This is due to the decreased polarity of the $p$-block metal-oxygen bond, which competes with the orthogonal configuration of the oxygen atomic orbitals involved in the bond. We find that the highly electronegative nature of the oxygen atom disrupts the vibrationless FCFs of $p$-block systems by causing large bond angle deflection upon excitation. This can be attributed to short metal-oxygen and oxygen-pseudohalogen bond lengths in the ground state, which induces bond-bond repulsion that works against the repulsive effects of the in-plane oxygen $3s$ lone pair and the T-shaped preference of the oxygen bonding orbitals. The result is an intermediate bond angle\footnote{The rotational \cite{apponi_millimeter-wave_1993} and photoionization \cite{pilgrim_photoionization_1993} spectra of AlOH suggests that the molecule is quasi-linear, with large amplitude bending motion. This is consistent with theoretical studies that indicate a flat ground state bending potential \cite{trabelsi_is_2018} which supports a true bent equilibrium at $\sim$160\degree \cite{vacek_x_1993} and a low-lying quasi-linear transition state \cite{li_characterization_2003}.} that is much larger than 90\degree, but smaller than 180\degree (see Fig. \ref{fig:fig4}(a)). As seen in our calculations (see Fig. \ref{fig:periodic}), the balancing of these competing effects creates a bond angle that is very sensitive to changes in the metal-oxygen bond hybridization, with $\Delta\theta>30^\circ$ shifts in the bond angle upon excitation from the ground state.

Using an atom larger than oxygen decreases the metal-ligand bond polarity, but also decreases bond repulsion. This produces a ground state bond angle that is close to 90$^\circ$ and also more stable upon excitation (see Fig. \ref{fig:fig4}(b)). We find that sulfur and selenium are ideal linker atoms that satisfy this requirement. Linker atoms heavier than period 4 (i.e. tellurium and polonium) also produce molecules with acceptable FCFs, although the larger spatial extent of the heavy chalcogen lone pair results in an acute ground state bond angle.  Fig. \ref{fig:periodic} displays a chart of ground and excited state bond angles and metal-ligand Mullikan charges against choice of linker atom.

Superficially, these results suggest a new, if counter-intuitive, heuristic: in bent, multivalent species, linker substitutions that create \textit{less polar} bonds may in fact result in \textit{more diagonal} FCFs. A more complete explanation is that laser coolable polyatomics with stable bond angles are likely to be found at opposite ends of the metal-ligand ionicity spectrum, where the molecule is either linear (i.e. CaOH) or T-shaped (i.e. CaSH), but not in the intermediate regime (i.e. AlOH), where there are multiple competing repulsive effects.
\subsection{Spin-orbit coupling and mass tuning}

The spin and spatial mechanisms that provide for highly diagonal cycling schemes in the multivalent OCCs considered in this work also lead to transitions that are partially forbidden by both spin and orbital selection rules. This makes the spontaneous emission rates, and therefore the performance of cycling schemes, strongly dependent on the strength of spin-orbit effects induced by the cycling center.

As discussed earlier, intersystem transitions, including the primary cycling lines for all three classes of multivalent polyatomics discussed in this paper, gain intensity from both dipole-allowed transitions and the ground state dipole moment via spin-orbit coupling. In operator form, the molecular spin-orbit term ($H_{so}\propto\mathbf{L}\cdot\mathbf{S}$) can be expressed as \cite{hirota_high-resolution_1985}:
\begin{align}
    H_{so}&=A_{so}[\mathbf{L}_z\mathbf{S}_z+(\mathbf{L}_+\mathbf{S}_-+
    \mathbf{L}_-\mathbf{S}_+)/2].\label{eq:Hso}
\end{align}
For valence electrons centered around a single atom, the molecular spin-orbit constant $A_{so}$ roughly scales as $Z^2$, thus favoring multivalent species with heavier OCCs, as seen earlier.  The $\mathbf{L}_z$ and $\mathbf{S}_z$ operators act diagonally on molecule-frame projections of orbital $|\Lambda\ket$ and spin $|\Sigma\ket$ angular momentum, respectively, while cross-terms in the remaining half of eq. \ref{eq:Hso} mix states with $\Delta\Lambda = \pm 1$ and $\Delta\Sigma =\mp 1$. MOs spaced apart by one quanta of orbital angular momentum are therefore maximally mixed by spin-orbit coupling. This leads to a geometric selection rule (colloquially referred to as the ``90 degree" or El-Sayed's rule by chemists) that orbitals with orthogonal spatial components have the largest amount of spin-orbit mixing and intersystem crossing \cite{salem_electronic_1972,shaik_qualitative_1978,doubleday_chemistry_1982,carlacci_spin-orbit_1987}. 

Unlike in linear and highly symmetric molecules, $\Lambda$ and $\Sigma$ are not good quantum numbers in ATMs due to the breakdown in the intermolecular axis of symmetry. Nonetheless, geometric selection rules for spin-orbit mixing still apply to component atomic orbitals, leading to rigorous mixing rules for MOs by their in-plane ($A'$) and out-of-plane ($A''$) character. Indeed, we observe in our candidate systems that strong SOC arises between electronic states of orthogonal symmetry, and weak or zero SOC manifests between electronic states of the same symmetry.

This has implications for expected scattering rates of intersystem cycling transitions in multivalent species, as the largest SOC -- and therefore strongest intensity borrowing -- is expected when there are nearby orthogonal-symmetry electronic states to which the spin-forbidden excited state can couple. Furthermore, the most intense dipole-allowed transitions occur between states of the same symmetry, which accordingly lead to more favorable intensities for nearby spin-forbidden lines. We can therefore deduce a general heuristic that more intense intersystem cycling transitions can be found in heavy multivalent species with level structure that supports: 1) an upper spin-forbidden electronic state with orthogonal symmetry to the ground state and 2) a closely lying dipole-allowed manifold with the same symmetry as the ground electronic state. 

This is the case for singlet group 13 and doublet group 14 systems considered in this paper, where an out-of-plane ($A''$) upper state couples strongly with an in-plane ($A'$) ground state and the dipole-allowed progression to a nearby in-plane ($A'$) excited state. Conversely, scattering rates in triplet group 15 systems are reduced due to the presence of a $^3A''$ ground state with the same symmetry as the $^1A''$ upper state as well as dominant intensity borrowing from a suppressed dipole transition moment between the $^3A''$ ground state and orthogonal $^3A'$ excited states.

\section{Outlook}\label{sec:outlook}

\subsection{Production}

Cryogenic buffer-gas cooling \cite{hutzler_buffer_2012} is a standard approach for producing large amounts ($>10^{10}$ per pulse) of cold, slow, gas-phase molecules. This technique enables rapid thermalization and relaxation of ``hot" reaction products through collisions with He (or other inert) buffer gas in a cryogenically cooled cell. A small aperture allows extraction of cold molecules via hydrodynamic entrainment in the buffer gas. A wide variety of organic and metallic molecules have been cooled using this technique, which is an important tool in molecular laser cooling and precision measurement. 

While many of our candidate molecules have yet to be experimentally produced and observed, cryogenically compatible pathways have been established for synthesizing numerous gas-phase metallic analogs at high densities. Typical approaches to gas-phase synthesis involve Nd:YAG laser ablation of a solid metal target, followed by the introduction of a gas-phase precursor via a heated capillary to produce the desired target species. For instance, thiol compounds with alkali \cite{kagi_laboratory_1997,janczyk_millimetersubmillimeter_2002,bucchino_trends_2013}, alkaline earth \cite{jarman_highresolution_1993, sheridan_high-resolution_2007, taleb-bendiab_millimeter-wave_2001, scurlock_molecular_1994}, (post-)transition metal centers \cite{okabayashi_microwave_2012, bucchino_examining_2017} -- including aluminum \cite{janczyk_laboratory_2006} -- (MSH) are routinely produced in the gas phase via evaporation or ablation of solid metals in the presence of H$_2$S gas. This method should be readily extendable to syntheses of MSH molecules with $p$-block metals. Synthesis of larger polyatomics could potentially involve using gas-phase methanethiol, (HSCH$_3$), ethanthiol (HSC$_2$H$_5$),  pentadienyl (C$_5$H$_5$), or benzene (C$_6$H$_6$) precursors, in a similar manner to the production of oxygen-containing polyatomics with capillary-introduced alcohol reactants (i.e. CH$_3$OH, C$_2$H$_6$O) \cite{paul_laser-induced_2019, kozyryev_determination_2019, crozet__2002, whitham_laser_1998, namiki_spectroscopic_1998, elhanine_laser_1998,paul_dispersed-fluorescence_2016, mitra_pathway_2022, zhu_functionalizing_2022}. Cryogenic yields can be further enhanced via state-selective excitation of reactants, as was demonstrated in the buffer-gas synthesis of YbOH molecules \cite{jadbabaie_enhanced_2020,pilgram_fine_2021}.

\subsection{Applications}

Multivalent optical cycling centers offer new avenues for quantum control, state preparation, and measurement with cold molecules. As mentioned in Sec. \ref{sec:intro}, multi-electron degrees of freedom have been leveraged extensively in cold atom experiments utilizing alkaline earth and transition metal species. Here, we have demonstrated a pathway to combining multivalent optical cycling centers with the unique and rich features of polyatomic molecular structure.

\subsubsection{Trapping and Control}
As in multi-electron atoms, the presence of two or more optically active electrons in molecular OCCs gives rise to tunable spin degrees of freedom, in both the ground and excited states of the molecule. This mechanism is generically unavailable in monovalent OCCs (e.g. AEM-pseudohalogens), which have a single optically active electron, and therefore photon cycling manifolds with only doublet spins and their half-integer $m_s=\{-1/2,+1/2\}$ magnetic projections.

Higher valence OCCs, on the other hand, can support states with a wide range of electronic spins $S>1/2$ and projections $m_s=\{-S, +S\}$. For instance, multivalent molecules with states possessing integer spins $S=\{0, 3\}$, such as those with group 13 and 15 OCCs, generically contain zero magnetic projection ($m_s=0$) states, where the sensitivities to external magnetic fields and couplings to internal hyperfine structure (e.g. $\mathbf{I}\cdot \mathbf{S}$) are suppressed, as well as high spin projection states $m_s=3$ where magnetic couplings are maximal. 

Naturally, multivalent OCCs also possess metastable electronic states with flipped spin multiplicity from the ground state. These states, which were discussed extensively in earlier sections in the context of photon cycling, could also serve as shelving states for state preparation and detection as well as long-lived storage or measurement states for quantum information and sensing applications. Crossings between the scalar polarizabilities of ground and spin-forbidden excited states, meanwhile, give rise to perturbation-free ``magic" dipole trapping wavelengths relative to transitions between the two states \cite{cooper_alkaline_2018, norcia_microscopic_2018, saskin_narrow-line_2019}.

Due to their high electron-spin states, multivalent molecules are compelling candidates for magnetically assisted slowing and trapping. One demonstrated approach is Zeeman-Sisyphus deceleration \cite{augenbraun_Zeeman-Sisyphus_2021, fitch_principles_2016} of a cryogenic beam, which can be followed by direct loading into a deep, superconducting ($\gtrsim$ 1 T) magnetic trap \cite{lu_magnetic_2014, campbell_magnetic_2007,weinstein_magnetic_1998} as well as subsequent transfer into an optical trap. This overcomes the limitation of direct Doppler slowing or trapping on narrowline intersystem transitions, while preserving photon budgets for high-fidelity state preparation and readout, or laser cooling of magnetically trapped molecules to ultralow Doppler temperatures. 

Alternatively, large radiative forces can be exerted directly on the molecules by using coherent techniques to bypass spontaneous emission. The use of multiphoton or stimulated optical techniques  -- such as CW polychromatic forces \cite{kozyryev_coherent_2018, metcalf_colloquium_2017, chieda_prospects_2011, aldridge_simulations_2016, galica_deflection_2018, wenz_large_2020} or ultrafast chirped $\pi$-pulses \cite{jayich_continuous_2014, long_suppressed_2019} -- for instance, could extend experimental flexibility by increasing effective scattering rates for narrow cycling transitions identified in this work. Strong field or light-dressing schemes, in analogy to optical quench techniques used to cool on narrowline transitions of light AEL atoms (e.g. Ca \cite{curtis_quenched_2001}, Mg \cite{rehbein_optical_2007}), may also be useful for decreasing effective lifetimes of excited cycling states in multivalent polyatomic species, potentially increasing scattering rates at the cost of mixing in transitions with less diagonal VBRs.

\subsubsection{Quantum simulation and information}
In addition to offering unique control properties, the internal structure generated by higher electronic valences in polyatomic molecules could offer new avenues for encoding quantum information, particularly in electron and nuclear spins \cite{rabl_hybrid_2006, park_second-scale_2017}, but also in low-lying rotational degrees of freedom. Structural asymmetry, as seen in bent MSH molecules, may also confer particular advantages for implementing error correction protocols \cite{albert_robust_2020}. Multivalent OCCs, in particular, possess fully tunable spin-orbital ($\mathbf{\Lambda}\cdot \mathbf{S}$) and spin-spin ($\mathbf{I}\cdot \mathbf{S}$) internal interactions via choice of $m_s$ state, where the electron spin angular momenta can be fully coupled or decoupled from other spin, orbital, and rotational modes. This could allow for selective spin-spin/spin-orbit interaction for information transfer or, alternatively, full decoupling for protection of encoded spins from decoherence.

Tunable spin couplings within multivalent molecules could also have utility in many-body quantum simulation. This can be illustrated via analogy with the structure of AEL atoms, which have been leveraged to simulate high-dimensional SU($N$) Hamiltonians through nuclear spin-independence \cite{gorshkov_two-orbital_2010,taie_su6_2012,hofrichter_direct_2016,ozawa_antiferromagnetic_2018,sonderhouse_thermodynamics_2020, schafer_tools_2020} as well as study multi-orbital physics via orbital Feshbach resonances  \cite{zhang_orbital_2015, zhang_controlling_2020} and spin-orbital effects \cite{galitski_spin-orbit_2013, livi_synthetic_2016, wall_synthetic_2016,kolkowitz_spinorbit-coupled_2017, bromley_dynamics_2018, zhang_controlling_2020}. While a detailed discussion on this subject is beyond the scope of this work, we posit that incorporating tunable multi-electron degrees of freedom with the unique benefits of polyatomic molecular structure \cite{kozyryev_precision_2017, yu_probing_2021} (e.g. high polarizability, metastable co-magnetometers) could point to new directions in studies of strongly correlated quantum systems, such as long-range lattice-spin models \cite{wall_simulating_2013, wall_quantum_2015, wall_realizing_2015}.

\subsubsection{Precision measurement}
The multivalent OCC paradigms developed in this work could also be useful for future precision measurements, particularly of fundamental symmetry violations~\cite{hutzler_polyatomic_2020,chupp_electric_2019,safronova_search_2018}. Extending optical cycling to molecules with multivalent centers not only offers access to optical clock metrology and new electronic degrees of freedom, but also unique \textit{nuclear} structure unavailable to earlier group elements. The heavy $p$-block elements Tl and Pb, for instance, are known to provide significant sensitivity to $T$- and $P$-violating effects such as the electron EDM and nuclear Schiff moments \cite{Baklanov_progress_2010, Kozlov_enhancement_2002, dzuba_electric_2002}. Inserting these heavy centers into polyatomic molecules would yield intrinsically sensitive internal states~\cite{isaev_laser-coolable_2017,kudrin_towards_2019} that simultaneously possess parity doublets and high polarizability useful for experimental measurements ~\cite{kozyryev_precision_2017,yu_probing_2021}.  While the molecule TlSH appears to have limited photon cycling capabilities, Tl- or Pb-containing polyatomic molecules functionalized with different ligands may yield better performance. The $p$-block also contains nuclei which are (or nearly are) doubly magic. This condition makes nuclear calculations of parity-violating effects, such as anapole moments, more tractable and provides a useful venue for interpretation of experimental results \cite{hergert_guided_2020, ronald_garcia_ruiz_private_nodate, jason_holt_private_nodate}. 

Similarly, it is worth exploring whether the approaches introduced in this work for engineering optical cycling into $p$-block centers could be extended to OCCs with even higher orbital angular momentum valence, such as transition metal centers with optically active $d$- and $f$- shells. Hg-containing molecules, for instance, are known to have extremely large sensitivity to $T,P$-violating effects \cite{Prasannaa_mercury_2015}. The triatomic HgOH~\cite{mitra_study_2021}, despite its high intrinsic sensitivity and polarizability, does not cycle photons for similar reasons to AlOH. Examining whether HgSH (and analogous molecules) may efficiently cycle photons is a well-motivated line of inquiry. 

We note that the new design principles described here could possibly be extended to molecular ions, which have powerful advantages for precision measurements~\cite{cairncross_precision_2017,yu_probing_2021,fan_optical_2021} and quantum science~\cite{chou_preparation_2017, patterson_method_2018}, but are challenging systems for incorporating optical cycling~\cite{ivanov_search_2020, oleynichenko_laser-coolable_2022, wojcik_prospects_2022}. The limitations of narrow linewidth transitions present in the types of molecules discussed here would be less relevant in ion traps, which do not rely on photon cycling to trap.

\subsubsection{Extensions to complex molecules}
In addition to exploring different choices of metal centers, ligand design may offer new internal structures and features. A particularly promising avenue is ligand functionalization of multivalent OCCs via an MS$R$-type motif where $R$ is a complex or chiral functional group. Prior experimental and theoretical studies with monovalent OCCs have found that AEM-pseudohalogen systems of the MO$R$, MS$R$, and M$R$-type possess properties favorable for laser cooling \cite{isaev_polyatomic_2016,paul_dispersed-fluorescence_2016, kozyryev_determination_2019, ivanov_towards_2019, augenbraun_molecular_2020, paul_electronic_2021, augenbraun_observation_2021}, including cases where $R$ is a complex organic ligand \cite{kozyryev_proposal_2016,ivanov_toward_2020, dickerson_franck-condon_2021, dickerson_optical_2021} up to as large as 60 atoms in size (i.e. fullerene) \cite{klos_prospects_2020}. Preliminary results suggest that large, multivalent M$S$R-type molecules also exhibit structural features conducive to photon cycling, including OCC-localized frontier orbitals with visible-wavelength energy spacings. Combining multi-electron OCC structure with complex electronic ligand degrees of freedom, such as via hypermetallic functionalization \cite{orourke_hypermetallic_2019, ivanov_two_2020}, or tuning Franck-Condon factors via ligand substitution \cite{dickerson_franck-condon_2021, ivanov_towards_2019,mitra_pathway_2022,zhu_functionalizing_2022} offers yet more unexplored design space. 

\section{Conclusion}
In this work, we have developed new design principles for engineering optical cycling into polyatomic molecules with multi-electron degrees of freedom, for which traditional design approaches fail. Using these paradigms, we have found several prototypical and candidate multivalent systems (M$X$H) that demonstrate properties favorable for optical cycling. Theoretical characterization indicates that these systems possess quasi-closed photon cycling schemes, exhibiting highly diagonal Franck-Condon factors and visible or near-visible transition wavelengths, with scattering rates dependent on spin-orbit mixing. These systems are prime candidates for further spectroscopic and computational investigations, which will be needed to devise tailored photon cycling and state control schemes for each molecule.

Through our analysis of multivalent OCCs, we have also elucidated the unique bonding and electrostatic mechanisms that enable highly diagonal cycling transitions in candidate systems. We have furthermore identified structural motifs that allow us to scale multivalent features to more complex polyatomic systems, including chiral and large organic functional groups.  Our results provide new directions towards designing optical cycling enters in polyatomic molecules with complex electronic structure.

\begin{acknowledgments} We thank Benjamin Augenbraun, Lan Cheng, Arian Jadbabaie, Anna Krylov, Nick Pilgram, and Pawe\l{} W\'{o}jcik for insightful discussions and feedback. P. Y. acknowledges support from the Eddleman Graduate Fellowship through the Institute for Quantum Information and Matter (IQIM), the Gordon and Betty Moore Foundation (7947), and the Alfred P. Sloan Foundation (G-2019-12502).  A. L. acknowledges support from the C. S.  Shastry Prize and the Caltech Associates SURF Fellowship. W. A. G. was supported by the Ferkel Chair. N. R. H. acknowledges support from the U.S. Department of Energy (DOE), Office of Science, Basic Energy Sciences (BES), under Award No. DE-SC0019245. The computations presented here were conducted in the Resnick High Performance Computing Center, a facility supported by Resnick Sustainability Institute at the California Institute of Technology. 
\end{acknowledgments}

\newpage
\appendix

\section{Details on vibronic structure}\label{appendix:vibronic}

\subsection{Molecular orbital configurations}

\begin{figure*}
	\includegraphics[width=2\columnwidth]{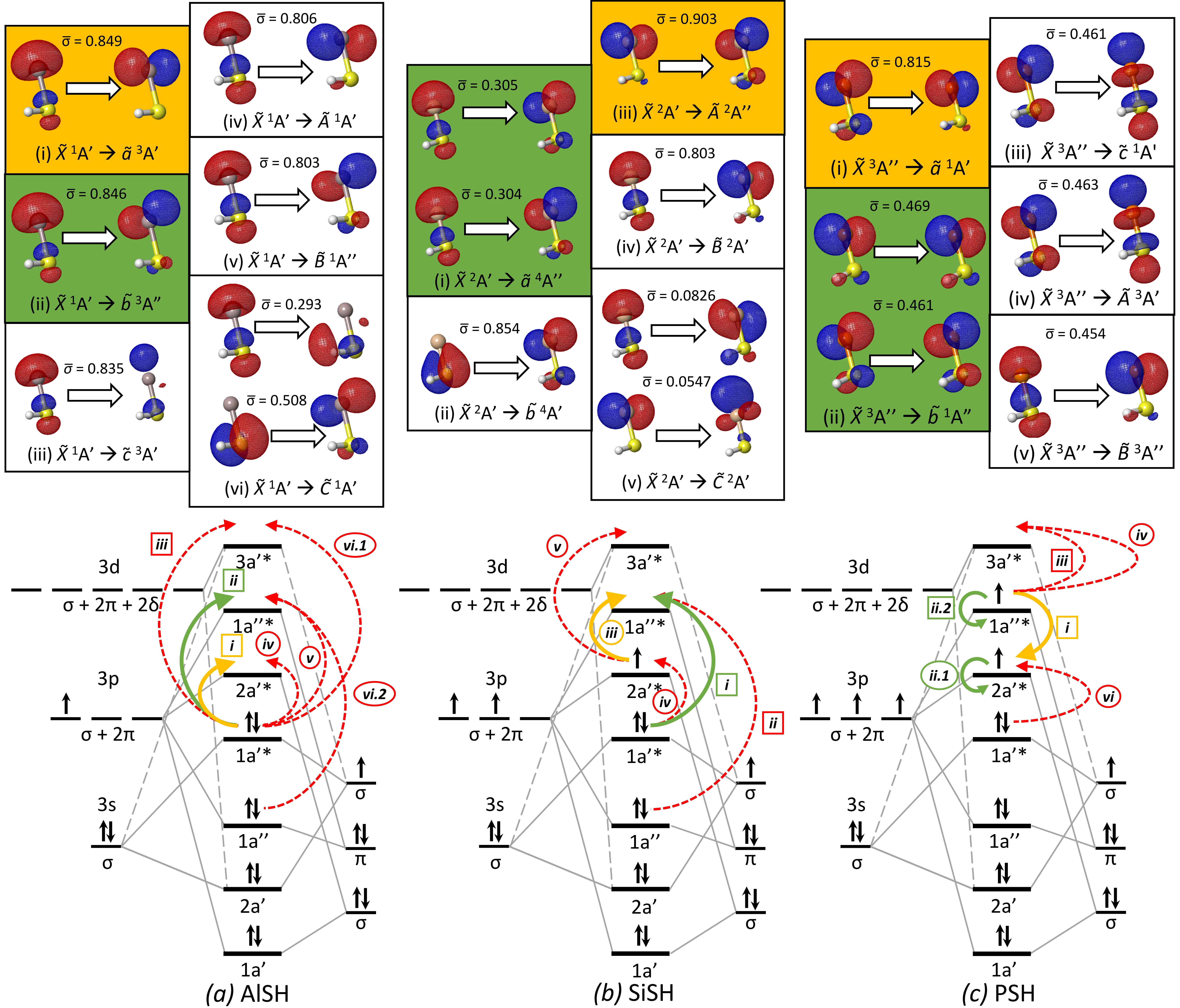}
	\caption{Natural transition orbitals (isovalue $=0.05$) for low-lying electronic transitions (top) and molecular orbital correlation diagrams (bottom) of three model multivalent species: (a) AlSH, (b) SiSH, and (c) PSH. Within each NTO panel are particle-hole pairs for the labeled transition, where $\bar{\sigma}$ denotes the relative NTO amplitude of each pair. Highlighted in green and yellow panels are the primary cycling and ground-to-intermediate state transitions, following the convention of table \ref{tab:FCF} in the main text. NTO calculations are performed using (a) EOM-EE-CCSD with a singlet ground state reference, (b) EOM-SF-CCSD with a quartet excited state reference, (c) and EOM-EE-CCSD with a singlet excited state reference. Below the NTO diagrams are MO correlation diagrams, which -- for each of the three molecules -- depict the ground state MO and spin configurations that are formed from the correlation of cycling center (left) and hydrosulfide ligand (right) orbitals. Each MO is labeled by its $C_s$ symmetry character ($A'$, $A''$), whereas the symmetry content of atomic and diatomic ligand orbital shells are labeled by $C_{\infty v}$ representations ($\sigma$, $\pi$, $\delta$...), which are described in the text. Curved arrows correspond to hole-particle NTO transitions depicted in the upper panels, which are individually identified by roman numerals, with subdivisions for transitions with multiple significant NTO components. Transitions labels surrounded by circles and squares indicate spin-allowed and spin-forbidden transitions, respectively. The cycling and ground-to-intermediate state transitions are identified with green and yellow solid arrows, following the earlier color convention, while the remaining transitions are identified with dashed red arrows.}
	\label{fig:nto}
\end{figure*}

The electronic structures of the MSH multivalent systems considered in this manuscript follow a common molecular orbital (MO) pattern. This can be seen by correlating the $3s$ and $3p$ shells of the cycling center (i.e. Al, Si, and P-like) to the valence orbitals of the hydrosulfide ligand (-SH), which are the S-H $\sigma$-bonding, S($3p\sigma$) unpaired electron, and S$(3p\pi)$ lone pair orbitals. This forms an MO progression that, in order of increasing energy, can be described as S-H $\sigma$-bonding ($1a'$), M-S $\sigma$-bonding ($2a'$), M-S $\pi$-bonding ($1a''$), M-S $\sigma$-antibonding ($1a'^*$), S-H $\sigma$-antibonding ($2a'^*$), and M-S $\pi$-antibonding ($1a''^*$) orbitals, as depicted in Fig. \ref{fig:nto}. A fourth non-bonding orbital ($3a'^*$) is formed from M(3$d\sigma$)-localized orbitals. These orbitals partially correlate to the ligand S($3p\sigma$) orbital and also mix into the M-S $\sigma$ bonding orbital, as indicated by the dotted line maps in the figure. Higher angular momentum orbitals out of the M($d$)-shell are not implicated in our analysis of photon cycling channels and therefore not examined for the sake of brevity.

For each model multivalent molecule (i.e. AlSH, SiSH, PSH), we perform natural transition orbital (NTO) analyses to elucidate the nature of the EOM-CC transitions calculated in this work. Hole-orbital isosurfaces (iso $= 0.05$) and amplitudes ($\bar{\sigma}$) for dominant NTO components are plotted in the upper panels of Fig. \ref{fig:nto} and numbered from (i) up to (vi), which correspond to unique EOM-CC transitions for each molecule. The EOM-CC calculations are performed as described in the main text:  AlSH using EOM-EE-CCSD with a singlet ground state reference, SiSH using EOM-SF-CCSD with a quartet excited state reference, and PSH using EOM-EE-CCSD with a singlet excited state reference. Curved lines with arrows overlaid on the MO schematics in Fig.\ref{fig:nto} depict the primary MO configurations involved in the NTO components for each transition. 

As discussed in Sec. \ref{sec:vibronic}, the highest occupied molecular orbital of the AlSH ground manifold corresponds to two paired electrons occupying the Al($s\sigma$) orbital or $1a'^*$, which is the M-S $\sigma$-antibonding orbital. Excited state triplet progressions, as depicted by the natural transition orbital analysis in \ref{fig:nto}(a), correspond to spin-forbidden excitations out of $1a'^*$ into the higher lying in-plane Al($3p$) or $2a'^*$ orbital [i: $\tilde{X}^1A'\to\tilde{a}^3A'$], which maps to the intermediate decay state, the out-of-plane Al($3p$) or $1a''^*$ orbital [ii: $\tilde{X}^1A'\to\tilde{b}^3A''$], which is the cycling line, and the in-plane Al($3d\sigma$) or $3a'^*$ orbital [iii: $\tilde{X}^1A'\to\tilde{c}^3A'$]. 

The ground singlet-to-excited singlet transitions are described by a similar progression, with the first two transitions corresponding to spin-preserving excitations from $1a'^*$ into the $2a'^*$ [iv: $\tilde{X}^1A'\to\tilde{A}^1A'$] and $1a''^*$ orbitals [v:  $\tilde{X}^1A'\to\tilde{B}^1A''$]. The third singlet-to-singlet transition [vi: $\tilde{X}^1A'\to\tilde{C}^1A'$] has mixed character, with NTO transitions from both the Al($s\sigma$) localized hole ($1a'^*$) exciting to a particle orbital with Al($d\sigma$) and S-H antibonding components [vi.1] -- which we approximately describe as $3a'^*$ -- as well as excitations from the subvalent Al-S $\pi$-bonding orbital ($1A''$) into the Al-S $\pi$-antibonding orbital ($1a''^*$) [vi.2]. The significant lengthing of the Al-S bond from upon excitation to both the $\tilde{c}^3A'$ and $\tilde{C}^1A'$ states confirms the charge-transfer nature of the NTO excitations to a delocalized Si($d\sigma$) orbital as depicted by [iii] and [vi.1].

For SiSH, the highest occupied molecular orbital (HOMO) corresponds to a single unpaired electron localized on the in-plane Si($3p\pi$) orbital ($2a'^*$). Excitations to the two lowest lying quartet states correspond to spin-forbidden transitions [i: $\tilde{X}^2A'\to\tilde{a}^4A''$] between the antibonding $1a'^*$ to the M-S $\pi$-antibonding orbital ($1a''^*$) and the M-S $\pi$-bonding orbital to the M-S $\pi$ antibonding orbital [ii: $\tilde{X}^2A'\to\tilde{b}^4A'$]. The proposed photon cycling channel for SiSH is between the ground and first excited quartet state, which have similar equilibrium geometries. Spin-allowed doublet-doublet excitations, meanwhile, correspond to transitions between the Si($3p\pi$)-localized $2a'^*$ orbitals to the M-S $\pi$-antibonding orbital ($1a''^*$) [iii: $\tilde{X}^2A'\to\tilde{A}^2A''$] and transitions between the Si($3s\sigma$) antibonding orbital ($1a'^*$) and the in-plane $\sigma$-antibonding orbital [iv: $\tilde{X}^2A'\to\tilde{B}^2A'$]. The third doublet-doublet transition [v: $\tilde{X}^2A'\to\tilde{C}^2A'$] has highly mixed character, which we approximately describe as excitations out of the in-plane Si($p\pi$) antibonding orbital ($2a'^*$) into the Si($d\sigma$) antibonding orbital ($3a'^*$).

As discussed earlier, PSH has a high-spin triplet ground state due to spin exchange and nodal plane effects. In the MO schematic, this configuration corresponds to two unpaired spins distributed between the in-plane ($2a'^*$) and out-of-plane ($1a''^*$) Si($3p\pi$)-localized antibonding orbitals. The first spin-forbidden transition [i: $\tilde{X}^3A''\to\tilde{a}^1A'$] the intermediate singlet state corresponds to a spin-forbidden from the upper Si($3p\pi$)-localized $1a''^*$ orbital to the lower $2a'^*$ orbital. This is a higher energy configuration due to the nodal separation between the two Si($p\pi$) lobes and the spin exchange energy, which is positive for paired electrons occupying the $2a'^*$ orbital. The cycling transition to the second singlet state [ii: $\tilde{X}^3A''\to\tilde{b}^1A''$], corresponds to a spin-forbidden within the upper $1a''^*$ orbital [ii.2], as indicated by the opposite phase convention in the hole-particle isosurfaces. A trivial $2a'^*\to2a'^*$ transition component [ii.1] is also identified by the NTO analysis, which we depict in Fig. \ref{fig:nto} for completeness. The third spin-forbidden transition [iii: $\tilde{X}^3A''\to\tilde{c}^1A'$] corresponds to spin-forbidden excitations from the unpaired Si($3\pi$)-localized $1a''^*$ antibonding orbital to the Si($3d\sigma$)-localized $3a'^*$ antibonding orbital.

The spin-preserving triplet-to-triplet transitions, meanwhile correspond to excitations from the frontier $1a''^*$ antibonding orbital to the Si($3d\sigma$)-localized $3a'^*$ antibonding orbital [iv: $\tilde{X}^3A''\to\tilde{A}^3A'$] and excitation of a single electron from the sub-valence Si($3s\sigma$)-localized $1a'^*$ antibonding orbital to pair to the unpaired electron in the lower Si($3p\pi$)-localized $2a'^*$ orbital [v: $\tilde{X}^3A''\to\tilde{B}^3A''$].
\FloatBarrier

\subsection{Valence bonding and repulsion}

\begin{figure}
	\includegraphics[width=0.95\columnwidth]{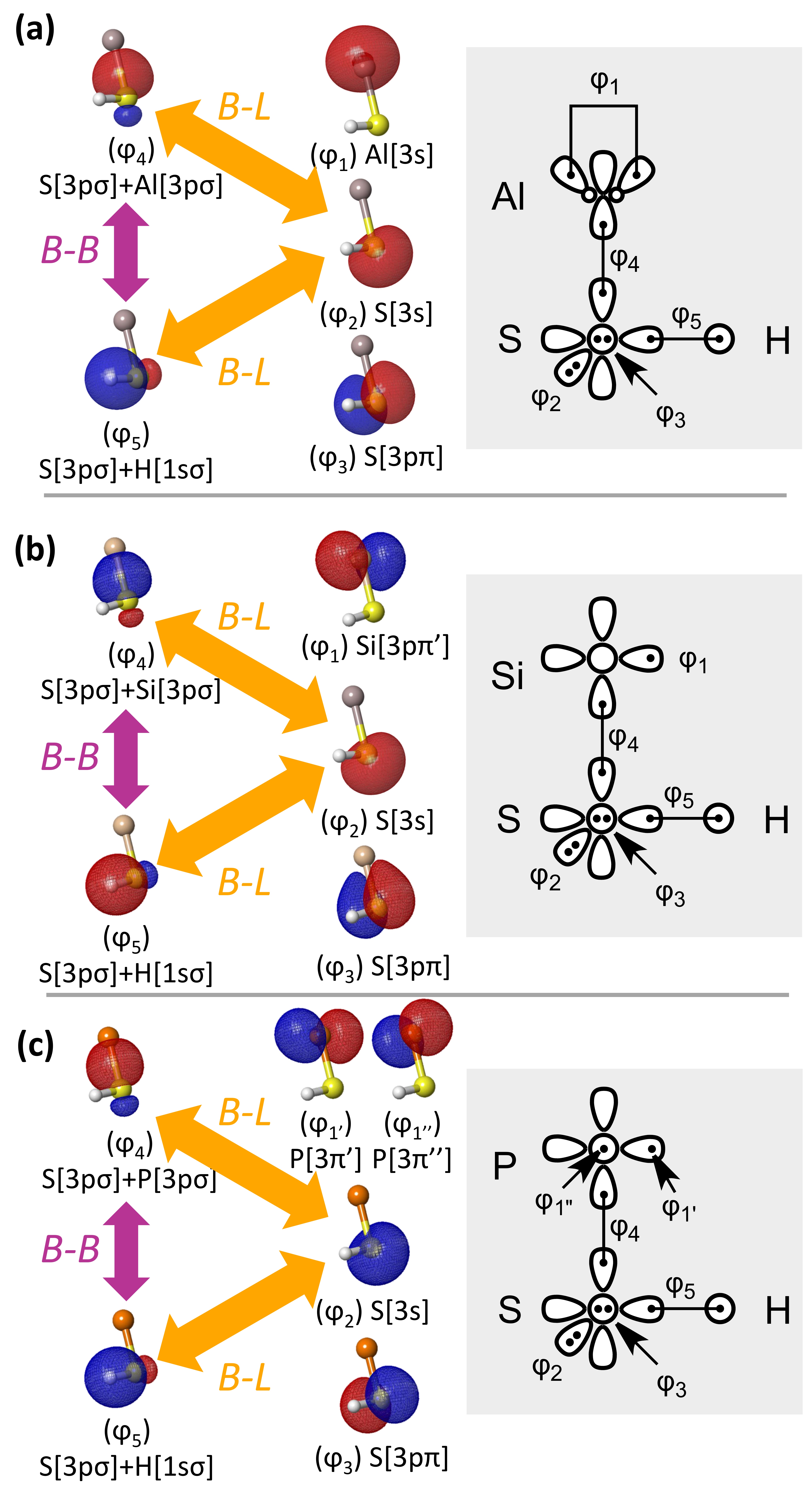}
	\caption{Generalized valence bond (GVB) natural orbitals and diagrams for molecules representative of the three multivalent classes: (a) AlSH, (b) SiSH, and (c) PSH. Depicted on the left side of each sub-panel are the five highest occupied bonding and non-bonding GVB natural orbitals (iso $= 0.05$), labeled $\phi_1$ to $\phi_5$ and by their main atomic orbital contributions. Details of the procedure used to generate the orbitals are described in Appendix \ref{appendix:vibronic}(2). The colored arrows $B-B$ (purple) and $B-L$ (yellow) indicate bond-bond and bond-lone pair repulsion between the plotted orbitals. The right inset of each sub-panel (gray background) contains the corresponding GVB diagram.}\label{fig:gvb}
\end{figure}

As discussed in Sec. \ref{sec:design}(a), the interplay of bonding, ionicity, and repulsion in MSH multivalent species plays an important role in determining the ground and excited state geometries, and by extension, the vibronic branching and viable photon cycling pathways. Bonding and repulsion are largely described by interactions between localized electron orbitals of the constituent atoms. The delocalized nature of the MO basis considered in the previous section therefore makes it unsuitable for visualizing these effects.

To examine the localized nature of bonding and lone pair orbitals depicted schematically in Fig. \ref{fig:fig4}, we generate generalized valence bond (GVB) natural orbitals \cite{goddard_description_1978, dunning_insights_2016} from canonical MOs using a Sano-type procedure \cite{sano_elementary_2000} for each of the multivalent classes. The Pipek-Mezey (PM) scheme \cite{pipek_fast_1989} is used to compute localized molecular orbitals that initialize the method, and all calculations are peformed over CCSD/aug-cc-pVTZ-optimized geometries. Fig. \ref{fig:gvb} depicts the localized GVB natural orbitals for representative systems from each of the three multivalent classes (AlSH, PSH, SiSH), as well as accompanying GVB diagrams. Plots and diagrams of valence bonding and non-bonding orbitals are labeled as $\phi_i$ using the index $i=1,..,5$. 

In the case of AlSH (see Fig. \ref{fig:periodic}(a)), $\phi_1$ denotes the paired Al-centered $3s\sigma$ orbital, which is the frontier GVB natural orbital and approximately describes the paired Al-centered cycling electrons. For the SiSH molecule (see Fig. \ref{fig:periodic}(b)), $\phi_1$ denotes the non-bonding Si-centered in-plane $3p\pi$ orbitals, which approximately describes the single unpaired Si-centered cycling electron. For the PSH molecule (see Fig. \ref{fig:periodic}(c)), $\phi_{1'}$ and $\phi_{1''}$ denote the unpaired in and out-of-plane $3p\pi$ orbitals that approximately describe the triplet unpaired P-centered cycling electrons.

The remaining four GVB natural orbitals have common descriptions among all three molecules. $\phi_2$ and $\phi_3$ refer to in-plane $s$ and out-of-plane $p\pi$ lone pairs that are centered on the S atom, respectively. $\phi_4$ and $\phi_5$, meanwhile, refer to the metal-S and S-H bonding orbitals, respectively. As discussed in Sec. \ref{sec:vibronic} and \ref{sec:design}, the bond angle of the molecule is determined by competition between the effects of bond-bond repulsion and lone pair-bond repulsion. Consistent with the notation in Fig. \ref{fig:fig4}, the labeled arrow $B-B$ (green) in Fig. \ref{fig:gvb} depicts the repulsive interaction between the metal-S ($\phi_4$) and S-H ($\phi_5$) bonding orbitals, while the $B-L$ arrows (yellow) depict the repulsion between the S in-plane lone pair ($\phi_2$) and the bonding orbitals ($\phi_4$) and ($\phi_5$).

\FloatBarrier
\begin{figure}
	\includegraphics[width=\columnwidth]{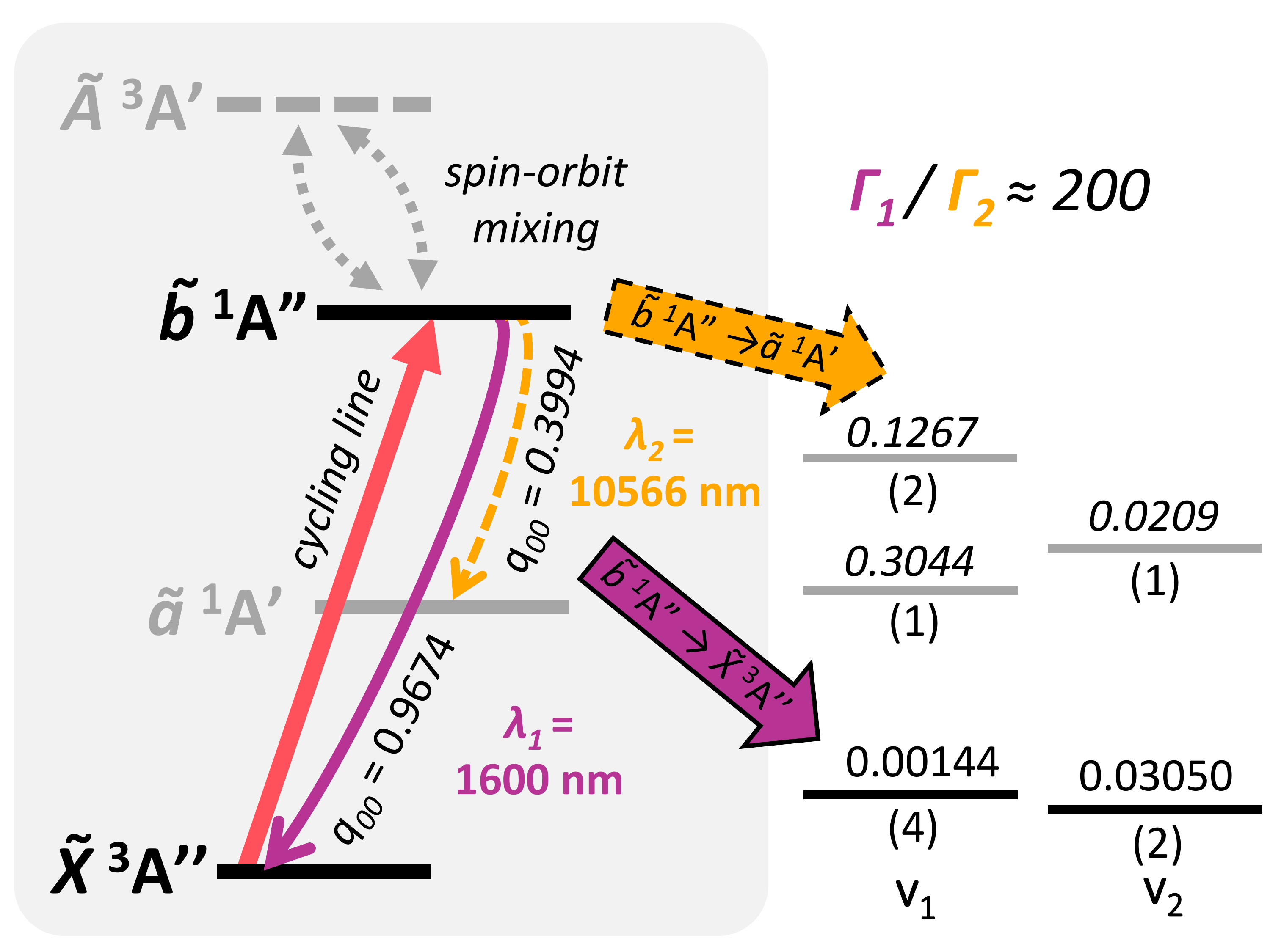}
	\caption{Optical cycling schemes for group 15 systems, demonstrated using BiSH vibrational branching ratios. The main excitation transition (red solid arrow) is from $\tilde X^3A''$ to $\tilde{b}^1A''$. On the left, solid curved (purple) and dashed curved (yellow) lines denote vibration-free decays to vibrational channels in the $\tilde X^3A''$ and $\tilde a^1A'$ manifolds, respectively, after spontaneous emission from the $\tilde{b}^1A''$ upper state. Pairs of gray dashed arrows depict the spin-orbit induced mixing between the first excited singlet state $\tilde{A}^3A'$ and the upper $\tilde{b}^1A''$ state.  Levels on the right hand side depict leading off-diagonal FCFs for decays to the $\tilde a^1A'$ and $\tilde X^3A''$ manifolds, with corresponding transition wavelengths denoted $\lambda_2$ and $\lambda_1$, respectively. Decimals above the levels denote the Franck-Condon factors (eq. \ref{eq:FCF}) normalized relative to the respective electronic transition, while numbers underneath indicate the vibrational quanta in each mode ($v_i$). Due to spin-orbit effects from the Bi center, more than 99.5$\%$ of decays out the $\tilde {b}^1{A''}$ state connect directly to the $\tilde{X}^3A''$ state, as indicated by the suppression factor ($\Gamma_1/\Gamma_2$) in the upper right hand corner (see Table \ref{tab:FCF}). Analogous pathways can be utilized to construct cycling schemes for other group 15 species, and branching patterns for other multivalent classes can be found in the Fig. \ref{fig:lc_scheme} in the main text and Fig. \ref{fig:SiSH_cycling} in the appendix. Level spacings are not drawn to scale.}
	\label{fig:BiSH_cycling}
\end{figure}

\subsection{Vibrational Closure Schemes}

\begin{figure}
	\includegraphics[width=\columnwidth]{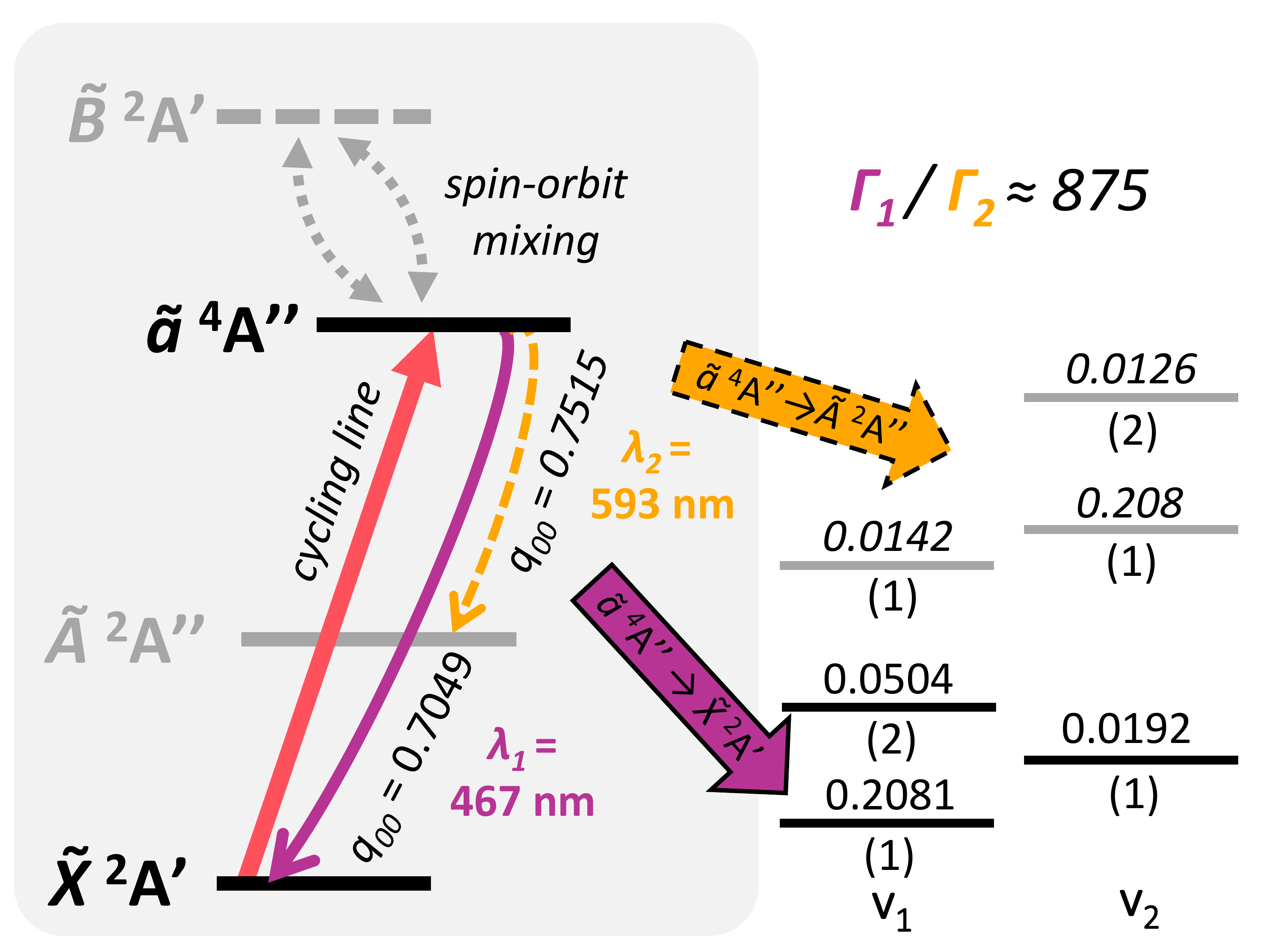}
	\caption{Optical cycling schemes for group 14 systems, demonstrated using SiSH branching ratios. The main excitation transition (red solid arrow) is from $\tilde X^2A'$ to $\tilde{a}^4A''$. On the left, solid curved (purple) and dashed curved (yellow) lines denote vibration-free decays to vibrational channels in the $\tilde X^2A'$ and $\tilde A^2A''$ manifolds, respectively, after spontaneous emission from the $\tilde{a}^4A''$ upper state. Pairs of gray dashed arrows depict the spin-orbit induced mixing between the first excited singlet state $\tilde{B}^2A'$ and the upper $\tilde{a}^4A''$ state.  Levels on the right hand side depict leading off-diagonal FCFs for decays to the $\tilde A^2A''$ and $\tilde X^2A'$ manifolds, with corresponding transition wavelengths denoted $\lambda_2$ and $\lambda_1$, respectively. Decimals above the levels denote the Franck-Condon factors (eq. \ref{eq:FCF}) normalized relative to the respective electronic transition, while numbers underneath indicate the vibrational quanta in each mode ($v_i$). Due to spin-orbit effects from the Si center, more than 99.88$\%$ of decays out the $\tilde {a}^4{A''}$ state connect directly to the $\tilde{X}^2A'$ state, as indicated by the suppression factor ($\Gamma_1/\Gamma_2$) in the upper right hand corner (see Table \ref{tab:FCF}). Analogous pathways can be utilized to construct cycling schemes for other group 14 species, and branching patterns for other multivalent classes can be found in the Fig. \ref{fig:lc_scheme} in the main text and Fig. \ref{fig:BiSH_cycling} in the appendix. Level spacings are not drawn to scale.}
	\label{fig:SiSH_cycling}
\end{figure}

In this section, we elaborate on the details of the optical cycling schemes for each class of multivalent molecules, as initially discussed in the main text in Sec. \ref{sec:pathways}, and depicted in Fig. \ref{fig:lc_scheme} for a model group 13 molecule, namely AlSH. Additional photon cycling and vibrational closure schemes -- for group 15 and 14 molecules -- shown here in Fig. \ref{fig:BiSH_cycling} and \ref{fig:SiSH_cycling}, respectively. Both figures depict the main cycling line (red arrows) between the ground state and spin-forbidden out-of-plane upper state, as well as dominant vibrational decays to off-diagonal vibrationals in the ground and intermediate electronic manifolds. 

In the case of class 15 molecules (Fig. \ref{fig:BiSH_cycling}), the ground state is a high-spin $\tilde{X}^3A''$ manifold, while the upper cycling state is a $\tilde{b}^1A''$ singlet state. This transition is allowed by spin-orbit induced intensity borrowing on dipole-allowed $\tilde{X}^3A''\to\tilde{A}^3A'$ channel. Dominant off-diagonal vibrational decays to both the ground and intermediate $\tilde{a}^1A'$ state are illustrated in blue and yellow (dotted boundary) panels, respectively. While the scheme here specifically utilizes FC data for BiSH (which has high electronic branching into the ground state), the photon cycling scheme depicted here is generally applicable to other group 15 systems listed in Table \ref{tab:FCF} of the main text.

Meanwhile, Fig. \ref{fig:SiSH_cycling} illustrates cycling and vibrational closure schemes for group 14 molecules, utilizing SiSH as a model system. Here, the ground state is a $\tilde{X}^2A'$ manifold, with a quasi-diagonal cycling transition to the upper $\tilde{a}^4A''$ state. This transition is allowed both by a combination of spin-orbit intensity borrowing from the $\tilde{X}^2A'\to \tilde{B}^2A'$ transition, as well as direct mixing of $\tilde{a}^4A''$ with the opposite-symmetry ground state. Leading vibrational decays to the ground and intermediate ($\tilde{A}^2A''$) are again depicted with an analogous color scheme to the figures for group 13 and 15 molecules. Note that although the vibrationless FCF of the intermediate $\tilde{a}^4A''\to\tilde{A}^2A''$ decay can be comparable, or sometimes even higher than the primary $\tilde{a}^4A''\to\tilde{X}^2A''$ line, it is not preferred as an optical cycling channel due to the relatively weak intensity of these transitions, as indicated in Table \ref{tab:FCF} of the main text, as well as the existence of rapid decays into the ground state from $\tilde{A}^2A''$ via relatively non-diagonal vibronic channels.

\subsection{Alternative ligands}

As discussed in Sec. \ref{sec:intro} and \ref{sec:vibronic} of the main text, traditional approaches to constructing laser coolable molecules by bonding cycling centers to highly electronegative ligands appear to fail for multivalent OCCs. This is evident from poor Franck-Condon performance for a variety of multivalent species built from aluminum centers bonded to common pseudohalogen ligands, including hydroxide (-OH), (iso)cynanide (-CN), cyanate (-NCO), acetylide (-CCH), fluoroacetylide (-CCF), and the superhalogen boron dioxide (-OBO). 

Data for these alternative systems, including low lying singlet and triplet Franck-Condon factors, Mulliken charge differences between metal and linker atoms, as well as calculated and experimental ligand electron affinities can be found in Table \ref{tab:ligand}.  All geometry, frequency, and charge density calculations are performed in line with methods described in the main text, using the EOM-EE-CCSD method and basis sets of aug-cc-pVTZ quality. Ligand electron affinity is calculated from the energy difference between an anion reference and a neutral configuration, with geometries optimized using CCSD and EOM-IP-CCSD, respectively.

Unlike aluminum hydroxide (AlOH), which adopts a bent geometry in the ground state, as described earlier, the additional pseudohalogen ligands we consider in this section and Table \ref{tab:ligand} bond to Al to form molecules that are predicted to have linear ground states, due to either (1) the use of an non-chalcogen linker atom (e.g. C, N) and therefore absence of lone-pair repulsive effects or (2) highly electronegative metal-ligand interactions that overcome lone-pair repulsive effects (e.g. AlOBO). Despite the apparent stability of their linear ground state configurations, all of these molecules also possess excited state geometries that are calculated to have either (1) significantly different metal-ligand bond lengths relative to the ground state and/or (2) large metal-ligand bend. Either of these properties leads to large off-diagonal FC behavior, which is fatal to achieving vibrationally closed photon cycling.

The precise nature of the bonding mechanisms that produce low vibrationless FCFs in the molecules (and transitions) considered here is non-generic and highly species-dependent. In addition, the set of alternative pseudohalogen ligands considered here -- while representative of molecules previously analyzed in the laser-cooling literature -- is non-exhaustive. We are nonetheless able to extract some useful observations.

First, there is no global correlation between bond ionicity, measured both in terms of ligand electron affinity and Mulliken charge difference between the linker and cycling atom, and Franck-Condon behavior, although slight improvements can be observed by switching to a more ionic isomer (e.g. AlCN vs. AlNC). Second, oxygen-containing ligands tend to perform poorly with multivalent cycling centers, regardless of ligand electronegativity. This is due to the interaction of the oxygen lone pair with high orbital angular momentum ($\Lambda > 0$) electrons centered on the Al metal. This is most easily seen with AlOH, as discussed earlier, but is also observed in AlNCO and AlOBO, where excitation from a linear Al$(3s\sigma)$-like ground state into Al$(3p\pi)$-like excited manifolds leads to bond angle bending and breaking of the linear symmetry due to interactions between the Al(3$p\pi$) orbitals and oxygen lone pairs. Third, despite the low performance of FCFs considered in Table \ref{tab:ligand}, spin-forbidden (triplet) transitions from the ground state do appear, for molecules considered, to perturb molecular geometries less relative to dipole-allowed singlet channels, which is consistent with earlier findings for AlSH.

\begin{table}[]
    \begin{ruledtabular}
    \begin{tabular}{c|c|c|c|c}
         Molecule & (a) $\tilde{A}\to\tilde{X}$ & (b) $\tilde{a}\to\tilde{X}$ & $\Delta Q_M$& EA (calc/exp) \\
        \hline
         AlCCH & 0.2533 
 &  0.4852
 & 0.0842 & 3.102/2.969(6)\textsuperscript{\cite{ervin_photoelectron_1991}}
 \\
         AlCCF & $<0.01$ & 0.48021
 & 0.106 & 3.383/$\gtrsim 3.4(8)$\textsuperscript{\cite{thynne_negative_1971}}
 \\
        AlNCO &  -- & $<0.01$ & 0.112 & 3.652/3.609(5)\textsuperscript{\cite{bradforth_photoelectron_1993}} \\
        AlCN & 0.1598
 & 0.3177 & 0.379 & 3.989/3.862(4)\textsuperscript{\cite{bradforth_photoelectron_1993}} \\
        AlNC & 0.4734 & 0.6417 & 0.387 & 3.989/3.862(4)\textsuperscript{\cite{bradforth_photoelectron_1993}} \\
         
        AlOBO & -- & -- & 1.041 & 4.465/4.46(3)\textsuperscript{\cite{zhai_vibrationally_2007}}\\
    \end{tabular}    
    \end{ruledtabular}\\
    \vspace{0.2 cm}
    \begin{ruledtabular}
    \begin{tabular}{c|c|c|c|c}
         Molecule & (c) $\tilde{B}\to\tilde{X}$ & (d) $\tilde{b}\to\tilde{X}$ & $\Delta Q_M$& EA (calc/exp)\\
        \hline
         AlSH & 0.2030 & 0.9974 & 0.609 & 2.322/2.317(2)\textsuperscript{\cite{breyer_high_1981}}\\
         AlOH & 0.0509 & 0.2468 & 0.981 & 1.761/1.82765(25)\textsuperscript{\cite{dewitt_high-resolution_2021}} 
    \end{tabular}
    \end{ruledtabular}
    \caption{Vibrationless  Franck-Condon factors ($q_{00}$) for electronic transitions labeled (a)-(d), metal-ligand Mulliken charge difference (a.u.), and ligand electron affinity (eV) for  (top) linear ($\Sigma^+$) ground state and (bottom)  bent ($A'$) ground state molecules. In order of appearance, the full term symbols for the transitions considered in the tables are: (a) $\tilde{A}^1\Pi\to \tilde{X}^1\Sigma^+$, (b) $\tilde{a}^3\Pi\to \tilde{X}^1\Sigma^+$, (c) $\tilde{B}^1A''\to \tilde{X}^1A'$, and (d) $\tilde{b}^3A''\to \tilde{X}^1A'$.  For the upper table, FCFs labeled by an asterisk indicate transitions between a nonlinear or quasilinear excited state and the linear ground state. Entries with ``$-$" indicate excited states that converge on strongly bent transition states due to the interaction of the Al$(3p\pi$) orbital with the oxygen lone pair. See Supplemental Materials for details.}
    \label{tab:ligand}
\end{table}

\section{Details on rotational structure and closure}\label{appendix:rotation}
\begin{figure*}
	\includegraphics[width=2\columnwidth]{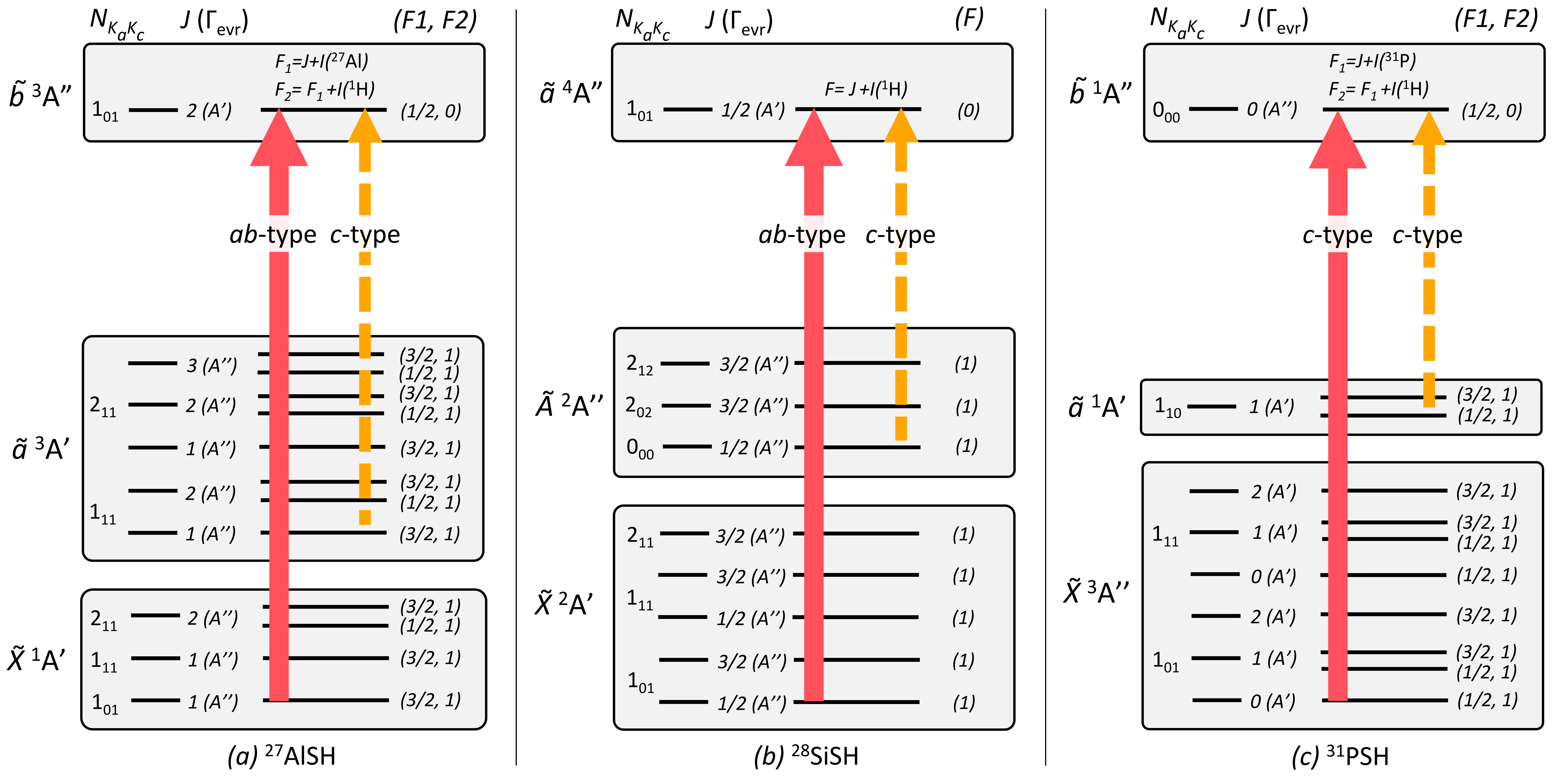}
	\caption{Proposed rotational repumping scheme based on optical transitions identified for (a) $^{27}$AlSH, (b) $^{28}$SiSH, and (c) $^{31}$PSH in Fig. \ref{fig:lc_scheme}. Levels are listed using $N_{K_a, K_c}$, $J$, and $F_i$ quantum numbers. In molecules with both metal $I(^Z\text{M})$ and hydrogen spins  $I(^1\text{H})$, $F_1=J+I(^Z\text{M})$ and $F_2=F_1+I(^1\text{H})$ denote the metal ($M$) and total hyperfine numbers, respectively, where $Z$ is the atomic mass of the metal. In open-shell states, the rotational levels are split by spin-orbit ($\mathbf{\Lambda_a} \cdot \mathbf{S}$) and spin-rotation ($\mathbf{N}\cdot \mathbf{S}$) couplings. For each spin-rotation branch, $\Gamma_\text{evr}$ denotes the combined rovibronic-spin symmetry $\Gamma_\text{el}\times\Gamma_\text{vib}\times\Gamma_\text{rot}$ of the level. The red, thick solid line denotes the primary $F\to F-1$ cycling transition from the ground state to the out-of-plane target state. The yellow, thin dashed lines denote rotationally closed repumps out of the intermediate manifold. Note that for spin-forbidden transition, only parity selection rules (see Table \ref{tab:selrule}) are applied, while for spin-allowed transitions, both angular momentum rules (see Table \ref{tab:rotclass}) and parity are applied.}
	\label{fig:rot_scheme}
\end{figure*}

Repumping of rotational decays in bent molecules can be achieved by addressing sidebands that are allowed by rotational and parity selection rules \cite{augenbraun_molecular_2020}. As discussed earlier, the broken symmetry of ATMs complicates the rotational level structure, which is described by the rigid rotor Hamiltonian
\begin{align}
    H_\text{rot}=A\mathbf{N}_a^2+B\mathbf{N}_b^2+C\mathbf{N}_c^2,
\end{align}
where we have three unique rotational constants ($A=\hbar/ I_a 4\pi$, $B=\hbar/I_b 4\pi$, $C=\hbar/I_c 4\pi$) and three spinless angular momentum operators $\mathbf{N}_i$ along the $i$-axis (see Fig. \ref{fig:fig3} for axis convention). The rigid rotor spectrum described by $H_\text{rot}$ is labeled by the term $N_{K_a, K_c}$. Here, $N$ is the total angular momentum  without electron $(S)$ and nuclear spin $(I)$, while $K_a$ and $K_c$ are approximate quantum numbers that correspond to the projection of $N$ onto the symmetry axis of the molecule if it were distorted to the limit of becoming a prolate ($a$-axis, $A>C$) or oblate ($c$-axis, $C>A$) symmetric top, respectively. Couplings to electronic and nuclear spins introduce the angular momenta $J=N+S$ and $F=J+I$. These quantum numbers encapsulate the level shifts induced by spin-orbit and hyperfine effects in the molecule.

\begin{table}[h]
\caption{\label{tab:rotclass}
Symmetries of $N_{K_a, K_b}$ rotational wavefunctions for prolate ATMs in terms of irreps of $C_{2v}$ and $C_{s}$ point groups, classified by the signs ($\sigma$) of $K_a$ and $K_c$. Parity selection rules for dipole transitions arise from the requirement $\Gamma_1\times\Gamma_2\subset\Gamma^*$, where \{$\Gamma_1$, $\Gamma_2$\} are the symmetries of two rovibronic states and $\Gamma^*$ is the electric dipole representation \cite{bunker_molecular_2006}.
}
\begin{ruledtabular}
\begin{tabular}{cccc}
 $\sigma(K_a)$&  $\sigma(K_c)$ & $\Gamma_\text{rot}(C_{2v})$ & $\Gamma_\text{rot}(C_s)$  \\
\hline
$+1$  & $+1$ & $A_1$ & $A'$    \\
$+1$ & $-1$  & $A_2$ & $A''$  \\
$-1$  & $-1$ & $B_1$ & $A''$  \\
$-1$ & $+1$ & $B_2$ & $A'$ 
\end{tabular}
\end{ruledtabular}
\end{table}

\begin{table}[h]
\caption{\label{tab:selrule}
Approximate angular momentum selection rules for dipole-allowed transitions in asymmetric top molecules, as adapted from \cite{augenbraun_molecular_2020}. These rules occur in addition to the standard rules on total angular momentum $\Delta J = 0$, $\pm 1$ and $J'\nleftrightarrow J'' = 0$, as well as the constraint that $K_a+K_c=N$ or $N+1$ for $0\leq K_{a,c} \leq N$. 
}
\begin{ruledtabular}
\begin{tabular}{cccc}
$\hat{\mu}$-axis & $\Delta K_a$ & $\Delta K_c$ & {Special cases} \\
\hline
$a$-type & 0 & $\pm 1$ & {$\Delta N\neq 0$ for $K'_a\to K_a''=0$} \\
$b$-type & $\pm 1$ & $\pm 1$ \\
$c$-type & $\pm 1$ & 0 & {$\Delta N\neq 0$ for $K'_c\to K_c''=0$}
\end{tabular}
\end{ruledtabular}
\end{table}
Each rotational state also has a symmetry classification ($\Gamma_\text{rot}$) in terms of the irreducible representations of the molecular symmetry (MS) group \cite{bunker_molecular_2006}. For bent ATMs with prolate rotational structure (i.e. $A>B>C$), the two choices of irreducible representations (irreps) in the $C_s$ symmetry group ($A'$, $A''$) map onto the parity ($+/-$) of $K_c$ for each rotational level, as seen in Table \ref{tab:rotclass}. Each choice of $|N,|K_a\ket$, for $|K_a|>0$ therefore forms a parity doublet that is closely spaced and split by the rotational asymmetry $\Delta E \propto [A(B+C)-(B+C)^2]/[4(2A-B-C)]$ \cite{polo_energy_1957}. Note that this is analogous to the hyperfine and centrifugally induced parity-doubling in symmetric top molecules \cite{nielsen_anomalous_1947,gunther-mohr_hyperfine_1954,watson_symmetry_1974,klemperer_can_1993}.

Combining rotational MS classifications with electronic and vibrational symmetries yields a total symmetry $(\Gamma_\text{evr})$, which is expressed as $\Gamma_\text{evr}=\Gamma_\text{el}\times\Gamma_\text{vib}\times\Gamma_\text{rot}$. This gives rise to dipole parity selection rules, which require that the product symmetry of two $E1$-connected rovibronic states ($\Gamma_1$,$\Gamma_2$) transform as the electric dipole representation ($\Gamma^*$).
\begin{align}
    \Gamma_1\times\Gamma_2\subset\Gamma^*
\end{align}
The representation $\Gamma^*$ is odd under inversion. For the two most common ATM symmetry groups $C_{2v}$ and $C_s$, this corresponds to $A_2$ and $A''$, respectively.

As summarized in Table \ref{tab:rotclass} and \ref{tab:selrule}, selection rules for dipole transitions can be constructed based on both the parity and the orientation of the electronic TDM ($\hat{\mu}$) relative to the rotational bands. It was recently shown \cite{augenbraun_molecular_2020} that these selection rules can be leveraged to realize rotationally closed cycling schemes in a broad class of monovalent, open-shell ATMs using two or fewer RF sidebands. We find that these principles can be readily extended for closing cycling transitions in multivalent closed-shell and high-spin systems, including mixed-character intercombination lines. 
\subsection{Singlet Ground States: Group 13}

For singlet group 13 molecules, spin-orbit mixing causes the $\tilde{X}^1A'\to\tilde{b}^3A''$ intercombination line to take on the character of the dipole-allowed $\tilde{X}^1A'\to\tilde{A}^1A'$ transition. This is a hybrid $ab$-type band, due to mixed alignment of the TDM with the two in-plane principal axes. The intermediate $\tilde{b}^3A''\to\tilde{a}^3A'$ decay, meanwhile, is a dipole-allowed $c$-type band, with the TDM aligned with the out-of-plane axis. This band follows $c$-type rotational selection rules.

The high-spin character of the nuclei of most group 13 isotopes requires tailored study of rotational and hyperfine branching to determine a suitable cycling scheme. To illustrate general principles, we perform an analysis given a $^{27}$Al cycling center, which has nuclear spin $I=5/2$. In open-shell and high-spin manifolds of asymmetric tops, spin-rotation (including spin-orbit) interactions split each $N_{K_a, K_c}$ rotational level into multiple $J=N+S$ branches. For the excited triplet manifolds of AlSH, each rotational level is split into three branches, $J=N-1, N,$ and $N+1$. The level diagrams in Fig. \ref{fig:rot_scheme} classify each rotational branch in terms of $\Gamma_\text{evr}$ for vibrationless cycling states. The hyperfine state including the $^{27}$Al spin $I(^{27}\text{Al})$ is denoted $F_1=J+I(^{27}\text{Al})$, while total hyperfine spin including hydrogen spin $I(^1\text{H})$ is denoted $F_2=F_1+I(^1\text{H})$.

Overlaid arrows in figure \ref{fig:rot_scheme} depict rotational closure schemes for vibrationless cycling and repump transitions in AlSH. The red, solid line depicts the main cycling transition between the $\tilde{X}^1A'$ ground state and the $\tilde{b}^3A''$ upper state. Due to the out-of-plane character of the excited state, the intercombination cycling line must be driven with even $\Delta K_c$ to respect parity, which are expected to override the (approximate) $ab$-type angular momentum selection rules in table \ref{tab:selrule} to yield $\Delta K_a = 0, \pm 1$ and $\Delta K_c = 0$ lines \cite{hougen_rotational_1964}. A rotationally closed cycling scheme for vibrationless and even-parity vibrational transitions can therefore be constructed by driving $F\to F-1$ (``type II") lines from $\tilde{X}^1A'$ $|1_{01},1, 3/2, 1\ket$, $|1_{11},1, 3/2, 1\ket$,  and $|2_{11},1, 3/2, 1\ket$ to $\tilde{b}^3A''$ $|1_{01},2, 1/2, 0\ket$, where we adopt the $|N_{K_a,K_c}, J, F_1, F_2\ket$ notation to describe angular angular momenta states. (Rotational schemes for repumping lines involving odd-parity vibrational transitions can be constructed using identical parity rules, while taking into account the symmetry of vibrational states ($\Gamma_\text{vib}$) involved.)

By contrast, for decays from the $\tilde{b}^3A''$ upper state to the $\tilde{a}^3A'$ state, the usual spin-allowed dipole selection rules apply. As a $c$-type transition, the allowed lines obey parity and $\Delta K_a = \pm 1$ and $\Delta K_c = 0$ angular momentum selection rules. A rotationally closed repumping scheme from the $\tilde{a}^3A'$ state is depicted via thin, yellow dashed lines in figure \ref{fig:rot_scheme}. A total of eight hyperfine sidebands are implicated in the repumping pathways. For both the cycling and repumping schemes, rotational spacings are expected to be on the GHz-scale, while hyperfine and spin-rotation splittings will typically be on the 10 MHz - 1 GHz scale, making individual sidebands fully addressable via optical modulation with a manageable number of seed lasers \cite{holland_synthesizing_2021}. Note that hyperfine-induced decays may populate other rotational states~\cite{norrgard_hyperfine_2017,fitch_laser-cooled_2021}

\subsection{Triplet Ground States: Group 15}

For triplet group 15 molecules, the $\tilde{X}^3A''\to \tilde{b}^1A''$ cycling line borrows intensity primarily from the $\tilde{X}^3A''\to \tilde{A}^3A'$ spin-allowed $c$-type transition. Fig. \ref{fig:rot_scheme} depicts the rotational closure scheme, including hyperfine effects, for the model system $^{31}$PSH. Both the $^{31}$P and proton nuclei have $I=1/2$, and we define two hyperfine quantum numbers as $F_1=J+I(^{31}\text{P})$ and $F_2=F_1+I(^{1}H)$.

Because both states involved in the $\tilde{X}^3A''\to \tilde{b}^1A''$ cycling line have the same electronic symmetry, parity rules require odd $\Delta K_c$ for vibrationless and even-parity vibrational transitions. We therefore expect that the allowed lines correspond to $\Delta K_a=0, \pm 1$ and $\Delta K_c = 0$ rules. In the case of $^{31}$PSH, driving from two ground rotational levels ($1_{01}$, $1_{11}$), each with four hyperfine sidebands ($|0, 1/2, 1\ket$, $|1, 1/2, 1\ket$, $|1, 3/2, 1\ket$, $|2, 3/2, 1\ket$) is sufficient to close the main cycling line to the upper $\tilde{b}^1A''|0_{00}, 0, 1/2, 0\ket$ state. Decays from $\tilde{b}^1A''|0_{00}, 0, 1/2, 0\ket$ to the intermediate $\tilde{a}^1A'$ state are spin-allowed and obey $c$-type selection rules. Closure of the intermediate decay can be achieved by repumping the $|1_{10}, 1, 1/2, 1\ket$ and $|1_{10}, 1, 3/2, 1\ket$ sidebands in the $\tilde{a}^1A'$ state.

\subsection{Doublet Ground States: Group 14} 
For doublet group 14 molecules, the $\tilde{X}^2A'\to \tilde{a}^4A''$ transition gains intensity primarily through spin-orbit mixing with the spin-allowed $\tilde{X}^2A'\to \tilde{B}^2A'$ and $\tilde{X}^2A'\to \tilde{C}^2A'$ $ab$-type transitions. The $\tilde{a}^4A''\to A^2A''$ intermediate decay channel, meanwhile, gains intensity primarily through mixing with the $\tilde{A}^2A''\to \tilde{B}^2A'$ $c$-type channel. As before, Fig. \ref{fig:rot_scheme} depicts the rotational closure scheme, accounting hyperfine splittings, for $^{28}$SiSH.  The lack of a high-spin triplet electronic manifold, as well as the spin-0 character of the $^{28}$Si nucleus is a simplifying factor, and we only define a single hyperfine number $F=J+I(^1\text{H})$ associated with the addition of hydrogen spin.

Due to their intercombination character, only parity selection and overall angular momentum selection rules are expected to apply for the cycling and intermediate decay lines in $^{28}$SiSH. For the cycling transition ($\tilde{X}^2A'\to \tilde{a}^4A''$), rotational closure (for parity-even vibrational transitions) can be achieved by driving type II transitions from three rotational states ($1_{01}, 1_{11}, 2_{11}$) and five spin-rotation components (with $F=1$) in $\tilde{X}^2A'$ to $\tilde{a}^4A''|1_{01}, 1/2, 0\ket$, where we denote the angular momentum states as $|N_{K_a, K_c}, J, F\ket$. The intermediate decay channel can similarly be repumped by addressing three rotational states ($0_{00}$, $2_{02}$, $2_{12}$) and three $F=1$ spin-rotation sidebands in $\tilde{A}^2A''$ to the $F=0$ upper state.

\begin{table*}[]
    \begin{minipage}{\columnwidth}
    (a) AlF
    \begin{ruledtabular}
    \begin{tabular}{cccc}
         $\tilde{X}^1\Sigma^+$ & Calc. & Exp. & Err.  \\
         \hline
         $r_0$& 1.674744 \AA &  1.65436(2) \AA  \textsuperscript{\cite{wyse_millimeter_1970}} & $+1.23\%$\\
         $\omega_0$ & 786.66 cm$^{-1}$ & 802.85(25) cm$^{-1}$ \textsuperscript{\cite{wyse_millimeter_1970}} & $-2.01\%$  \\
         $d_0$ & $1.5321$ D & $1.515(4)$ D \textsuperscript{\cite{truppe_spectroscopic_2019}} & $+1.12\%$\\ \\
         $\tilde{a}^3\Pi$ \\
         \hline
         $r_0$ & 1.6687 \AA & 1.6476 \AA  \textsuperscript{\cite{herzberg_molecular_1979}} & $+1.28\%$\\
         $\omega_0$ & 812.89 cm$^{-1}$ & 830.3(3) cm$^{-1}$ \textsuperscript{\cite{kopp_rotational_1976}} & $-2.10 \%$\\
         $\tau$ & $13.56$ ms & $> 1$ ms \textsuperscript{\cite{truppe_spectroscopic_2019}} & $-$ \\
         $d_0$ & 1.8476 D & 1.780(3) D \textsuperscript{\cite{truppe_spectroscopic_2019}} & $+3.78\%$ \\
         $T_0$ & 26320.276 cm$^{-1}$ & 27255.15(2) cm$^{-1}$ \textsuperscript{\cite{truppe_spectroscopic_2019}} & $-3.43\%$ \\ \\
         $\tilde{A}^1\Pi$ & \\
         \hline
         $r_0$ &  1.6680 \AA & 1.6485 \AA  \textsuperscript{\cite{herzberg_molecular_1979}} & $+1.95\%$\\
         $\omega_0$ & 791.52 cm$^{-1}$ & 803.9(5) cm$^{-1}$ \textsuperscript{\cite{rowlinson_absorption_1953}} & $-1.54\%$ \\
         $\tau$ & $1.873$ ns & $1.90(3)$ ns  \textsuperscript{\cite{truppe_spectroscopic_2019}} & $-1.42\%$ \\
         $d_0$ & 1.4609 D & 1.45(2) D \textsuperscript{\cite{truppe_spectroscopic_2019}} & $+0.75\%$  \\
         $T_0$ & 43875.173 cm$^{-1}$ & 43950.285(10) cm$^{-1}$ \textsuperscript{\cite{truppe_spectroscopic_2019}} & $-0.17\%$ 
    \end{tabular}
    \end{ruledtabular}
    
    \vspace{0.25 cm}
    (c) AlSH
    \begin{ruledtabular}
    \begin{tabular}{cccc}
         $\tilde{X}^1A'$ & Calc. & Exp. & Err.  \\
         \hline
         $r_0$(Al-S)& 2.2646 \AA &  2.240(6) \AA  \textsuperscript{\cite{janczyk_laboratory_2006}} & $+1.07\%$\\
         $r_0$(S-H)& 1.3464 \AA &  1.36(4) \AA  \textsuperscript{\cite{janczyk_laboratory_2006}} & $-1.00\%$\\
         $\angle$(Al-S-H)& 90.19\degree & 88.5$\pm$5.8\degree  \textsuperscript{\cite{janczyk_laboratory_2006}}& $+1.91\%$ \\
    \end{tabular}
    \end{ruledtabular}
    \end{minipage}
    \quad
    \begin{minipage}{\columnwidth}
    (b) AlOH
    \begin{ruledtabular}
    \begin{tabular}{cccc}
         $\tilde{X}^1A'$ & Calc. & Exp. & Err.  \\
         \hline
         $r_0$(Al-O)& 1.6924 \AA &  1.682 \AA  \textsuperscript{\cite{apponi_millimeter-wave_1993}} & $+0.62\%$\\
         $r_0$(O-H)& 0.9498 \AA &  0.878 \AA  \textsuperscript{\cite{apponi_millimeter-wave_1993}, \footnote{The experimentally determined O-H bond length is unusually short, which is suggested in \cite{apponi_millimeter-wave_1993} to be an artifact of large amplitude vibrations in the observed states.}} & $+8.18\%$\\
         $\angle$(Al-O-H)& 157.54 \AA &  $\sim 160$\degree  \textsuperscript{\cite{apponi_millimeter-wave_1993}}\textsuperscript{, }\footnote{Inferred in \cite{apponi_millimeter-wave_1993} from experimentally determined bond lengths and comparisons against theoretical predictions \cite{vacek_x_1993, trabelsi_is_2018} of a bent ground state with a low barrier to quasi-linearity \cite{li_characterization_2003}. This is consistent with rotational \cite{apponi_millimeter-wave_1993} and photonionization spectra \cite{pilgrim_photoionization_1993}, which suggest that AlOH is, on average, a near-linear molecule with large-amplitude zero point bending.} & $-$\\
         $\omega$(Al-O) & 834.10 cm$^{-1}$ & 810.7 cm$^{-1}$ \textsuperscript{\cite{wang_infrared_2007}} & $+2.89\%$  \\
         $\omega$(O-H) & 4035.31 cm$^{-1}$ & 3787.0 cm$^{-1}$ \textsuperscript{\cite{wang_infrared_2007}} & $+6.56\%$  \\ \\
         $\tilde{A}^1A'$ \\
         \hline
         $\omega$(Al-O) & 743.36 cm$^{-1}$ & 807.6 cm$^{-1}$ \textsuperscript{\cite{pilgrim_photoionization_1993}} & $-7.95\%$  \\ 
         $\omega$(O-H) & 3867.45 cm$^{-1}$ & 3258.4 cm$^{-1}$ \textsuperscript{\cite{pilgrim_photoionization_1993}} & $+18.7\%$  \\
         $\omega$(bend) & 624.80 cm$^{-1}$ & 636.6 cm$^{-1}$ \textsuperscript{\cite{pilgrim_photoionization_1993}} & $-1.85\%$  \\ 
         $T_0$ & 39810 cm$^{-1}$ & 40073 cm$^{-1}$ \textsuperscript{\cite{pilgrim_photoionization_1993}} & $-0.65\%$ \\ \\
         $\tilde{B}^1A''$ & \\
         \hline
         $\omega$(Al-O) & 747.55 cm$^{-1}$ & 760.3 cm$^{-1}$ \textsuperscript{\cite{pilgrim_photoionization_1993}} & $-1.68\%$  \\ 
         $\omega$(bend) & 598.92 cm$^{-1}$ & 581.0 cm$^{-1}$ \textsuperscript{\cite{pilgrim_photoionization_1993}} & $+3.08\%$  \\ 
         $T_0$ & 41550 cm$^{-1}$ & 41747 cm$^{-1}$ \textsuperscript{\cite{pilgrim_photoionization_1993}} & $-0.47\%$ 
    \end{tabular}
    \end{ruledtabular}
    \end{minipage}
    
    \caption{Calculated and experimental molecular constants for (a) AlF, (b) AlOH, and (c) AlSH. As described in the main text, ab initio values are calculated using the EOM-EE-CCSD method with aug-cc-pVTZ basis sets. Spin-orbit data (for estimating lifetimes of metastable states) is computed perturbatively using a Breit-Pauli Hamiltonian.}
    \label{tab:benchmark}
\end{table*}

\bibliography{ref}

\FloatBarrier

\setcounter{section}{0}
\setcounter{equation}{0}
\setcounter{figure}{0}
\setcounter{table}{0}
\makeatletter
\renewcommand{\theequation}{S\arabic{equation}}
\renewcommand{\thefigure}{S\arabic{figure}}
\renewcommand{\thetable}{S\arabic{table}}
\renewcommand{\thesection}{S\arabic{section}}

\begin{table*}
\caption{\label{tab:geos}
Electronic energies (eV), rovibrational constants (cm$^{-1}$), and geometries (\AA, degrees) for group 13 MSH molecules considered in Section II. 
}
\begin{ruledtabular}
\begin{tabular}{lllllllllll}
BSH & $\Delta E$ & $r$(B-S) & $r$(S-H) & $\angle$(B-S-H) & A & B & C & $v_1$ & $v_2$ & $v_3$\\
\hline
$\tilde{X}^1A'$ & -- & 1.822 & 1.348 & 87.65 & 9.492 & 0.616 & 0.579 & 587.07 & 774.14 & 2676.45\\
$\tilde{a}^3A'$ & 1.461 & 1.736 & 1.357 & 102.50 & 9.981 & 0.668 & 0.626 & 845.59 & 936.26 & 2534.03\\
$\tilde{b}^3A''$ & 2.396 & 1.837 & 1.347 & 89.21 & 9.512 & 0.605 & 0.569 & 464.17 & 757.76 & 2693.52\\
$\tilde{A}^1A'$ & 3.694 & 1.705 & 1.407 & 105.90 & 9.623 & 0.687 & 0.641 & 815.13 & 939.58 & 1935.49\\
$\tilde{B}^1A''$ \footnote[1]{State is unbound and pre-dissociative. Geometry and energies correspond to saddlepoints on the potential energy surface.} & 4.122 & 1.871 & 1.348 & 88.60 & 9.501 & 0.584 & 0.550 & 783.63$i$ & 496.84 & 2049.82\\
$\tilde{c}^3A'$ & 4.759 & 1.892 & 1.829 & 179.84 & -- & 0.4870 & -- & 187.75 & 565.26 & 801.93\\
$\tilde{C}^1A'$ & 4.918 & 2.563 & 1.347 & 86.48 & 9.535 & 0.311 & 0.302 & 297.31 & 336.29 & 2692.98\\
\hline
AlSH & $\Delta E$ & $r$(Al-S) & $r$(S-H) & $\angle$(Al-S-H) & A & B & C & $v_1$ & $v_2$ & $v_3$\\
\hline
$\tilde{X}^1A'$ & -- & 2.264 & 1.346 & 90.19 & 9.523 & 0.221 & 0.216 & 426.18 & 496.56 & 2684.52\\
$\tilde{a}^3A'$ & 2.340 & 2.203 & 1.344 & 98.59 & 9.833 & 0.233 & 0.227 & 428.68 & 502.46 & 2701.20\\
$\tilde{b}^3A''$ & 2.744 & 2.259 & 1.346 & 90.09 & 9.522 & 0.222 & 0.217 & 429.62 & 497.47 & 2684.65\\
$\tilde{A}^1A'$ & 3.694 & 2.260 & 1.351 & 100.29 & 9.835 & 0.216 & 0.221 & 433.35 & 486.85 & 2649.27\\
$\tilde{B}^1A''$ & 4.097 & 2.404 & 1.347 & 89.30 & 9.514 & 0.193 & 0.197 & 325.03 & 497.52 & 2681.22\\
$\tilde{c}^3A'$ & 4.161 & 2.899 & 1.347 & 89.88 & 9.512 & 0.135 & 0.133 & 17.48 & 478.35 & 2677.03\\
$\tilde{C}^1A'$ \footnotemark & 4.381 & 3.153 & 1.346 & 90.27 & 9.520 & 0.114 & 0.113 & 107.77$i$ & 457.08 & 2679.83\\
\hline
GaSH & $\Delta E$ & $r$(Ga-S) & $r$(S-H) & $\angle$(Ga-S-H) & A & B & C & $v_1$ & $v_2$ & $v_3$\\
\hline
$\tilde{X}^1A'$ & -- & 2.293 & 1.345 & 90.21 & 9.539 & 0.144 & 0.142 & 346.65 & 490.68 & 2691.94\\
$\tilde{a}^3A'$ & 2.699 & 2.207 & 1.345 & 98.98 & 9.835 & 0.154 & 0.152 & 375.87 & 701.43 & 2664.27\\
$\tilde{b}^3A''$ & 3.113 & 2.283 & 1.346 & 89.66 & 9.531 & 0.145 & 0.143 & 327.39 & 616.18 & 2694.97\\
$\tilde{B}^1A''$ & 3.969 & 2.819 & 1.348 & 105.57 & 10.316 & 0.0943 & 0.0934 & 147.42 & 279.02 & 2655.63\\
$\tilde{c}^3A'$ & 3.999 & 3.542 & 1.344 & 179.99 & -- & 0.0585 & -- & 61.57 & 338.55 & 2712.45\\
$\tilde{C}^1A'$ & 4.167 & 3.152 & 1.347 & 88.18 & 9.517 & 0.144 & 0.142 & 113.60 & 201.60 & 2692.19\\
\hline
InSH & $\Delta E$ & $r$(In-S) & $r$(S-H) & $\angle$(In-S-H) & A & B & C & $v_1$ & $v_2$ & $v_3$\\
\hline
$\tilde{X}^1A'$ & -- & 2.503 & 1.345 & 91.06 & 9.549 & 0.105 & 0.104 & 300.29 & 472.56 & 2691.19 \\
$\tilde{a}^3A'$ & 2.616 & 2.423 & 1.343 & 96.96 & 9.753 & 0.112 & 0.110 & 316.66 & 642.26 & 2698.88\\
$\tilde{b}^3A''$ & 2.921 & 2.485 & 1.345 & 90.11 & 9.538 & 0.106 & 0.105 & 283.35 & 594.64 & 2696.07\\
$\tilde{A}^1A'$ & 3.556 & 2.602 & 1.345 & 96.08 & 9.675 & 0.0968 & 0.0959 & 160.14 & 530.21 & 2684.08\\
$\tilde{c}^3A'$ & 3.719 & 3.854 & 1.343 & 179.98 & -- & 0.0432 & -- & 33.38 & 367.44 & 2717.21\\
$\tilde{B}^1A''$ & 3.745 & 3.132 & 1.346 & 87.95 & 9.535 & 0.0671 & 0.0667 & 33.76 & 257.11 & 2702.37\\
$\tilde{C}^1A'$ & 3.885 & 3.336 & 1.347 & 89.94 & 9.516 & 0.0591 & 0.0587 & 67.55 & 136.48 & 2692.06\\
\hline
TlSH & $\Delta E$ & $r$(Tl-S) & $r$(S-H) & $\angle$(Tl-S-H) & A & B & C & $v_1$ & $v_2$ & $v_3$\\
\hline
$\tilde{X}^1A'$ & -- & 2.576 & 1.345 & 91.08 & 9.554 & 0.0893 & 0.0885 & 272.59 & 454.01 & 2692.25\\
$\tilde{a}^3A'$ & 3.100 & 2.541 & 1.344 & 97.10 & 9.740 & 0.0915 & 0.0907 & 212.35 & 560.94 & 2691.48\\
$\tilde{b}^3A''$ & 3.330 & 2.661 & 1.345 & 88.65 & 9.536 & 0.0838 & 0.0831 & 157.92 & 510.66 & 2701.71\\
$\tilde{A}^1A'$ & 3.512 & 2.968 & 1.345 & 95.09 & 9.635 & 0.0672 & 0.0667 & 71.31 & 275.75 & 2699.78\\
$\tilde{c}^3A'$ & 3.538 & 3.170 & 1.347 & 86.88 & 9.527 & 0.0591 & 0.0587 & 106.87 & 197.40 & 2692.33\\
$\tilde{B}^1A''$ & 3.563 & 3.103 & 1.346 & 105.11 & 10.280 & 0.0612 & 0.0608 & 118.94 & 202.09 & 2678.72\\
$\tilde{C}^1A'$ & 3.704 & 3.388 & 1.347 & 89.60 & 9.515 & 0.0517 & 0.0514 & 87.27 & 162.82 & 2692.33\\
\end{tabular}

\end{ruledtabular}
\end{table*}

\begin{table*}
\caption{\label{tab:geos3}
Electronic energies (eV), rovibrational constants (cm$^{-1}$), and geometries (\AA, degrees) for group 14 MSH molecules considered in Section II. 
}
\begin{ruledtabular}
\begin{tabular}{lllllllllll}
CSH & $\Delta E$ & $r$(C-S) & $r$(S-H) & $\angle$(C-S-H) & A & B & C & $v_1$ & $v_2$ & $v_3$\\
\hline
$\tilde{X}^2A'$ & -- & 1.664 & 1.361 & 101.64 & 9.855 & 0.637 & 0.682 & 805.80 & 921.01 & 2512.34\\
$\tilde{A}^2A''$ & 1.085 & 1.767 & 1.351 & 85.21 & 9.488 & 0.616 & 0.578 & 586.73 & 774.00 & 2662.54\\
$\tilde{a}^4A''$ & 2.121 & 1.720 & 1.347 & 97.67 & 9.761 & 0.602 & 0.642 & 802.76 & 942.57 & 2650.00\\
$\tilde{B}^2A'$ & 3.488 & 1.352 & 1.755 & 98.47 & 9.746 & 0.616 & 0.579 & 729.04 & 930.90 & 2627.00\\
$\tilde{C}^2A'$ & 4.380 & 1.772 & 1.425& 114.79 & 10.679 &  0.589 & 0.558 & 236.54 & 955.64 & 2031.77\\
\hline
SiSH & $\Delta E$ & $r$(Si-S) & $r$(S-H) & $\angle$(Si-S-H) & A & B & C & $v_1$ & $v_2$ & $v_3$\\
\hline
$\tilde{X}^2A'$ & -- & 2.128& 1.346 & 99.73 & 9.865 & 0.243 & 0.237 & 516.11 & 681.41 & 2667.08\\
$\tilde{A}^2A''$ & 0.593  & 2.197 & 1.347 & 88.20 & 9.497 & 0.230 & 0.225 & 466.37 & 566.04 & 2675.51\\
$\tilde{a}^4A''$ & 2.646 & 2.187 & 1.344 & 96.01 & 9.705 & 0.226 & 0.232 & 434.40 & 739.17 & 2696.67 \\
$\tilde{B}^2A'$ & 3.793 & 2.120 & 1.355 & 103.01 & 10.003 & 0.245 & 0.239 & 417.60 & 827.97 & 2556.84 \\
$\tilde{C}^2A'$ & 3.894 & 2.640 & 1.349 & 103.33 & 10.103 & 0.158 & 0.156 & 263.62 & 363.54 & 2673.15\\
\hline
GeSH & $\Delta E$ & $r$(Ge-S) & $r$(S-H) & $\angle$(Ge-S-H) & A & B & C & $v_1$ & $v_2$ & $v_3$\\
\hline
$\tilde{X}^2A'$ & -- & 2.209& 1.345 & 98.69 & 9.830 & 0.148 & 0.150 & 410.73 & 654.91 & 2680.73\\
$\tilde{A}^2A''$ & 0.525  & 2.271 & 1.347 & 89.28 & 9.513 & 0.141 & 0.143 & 384.28 & 557.66 & 2681.62\\
$\tilde{a}^4A''$ & 2.845 & 2.324 & 1.344 & 95.86 & 9.690 & 0.134 & 0.136 & 242.58 & 649.39 & 2700.67\\
$\tilde{B}^2A'$ & 3.042 & 3.019 & 1.345 & 93.50 & 9.595 & 0.0810 & 0.0803 & 75.02 & 272.28 & 2707.13\\
$\tilde{C}^2A'$ & 3.670 & 2.755 & 1.347 & 101.41 & 9.963 & 0.0968 & 0.0958 & 146.72 & 350.71 & 2687.71\\
\end{tabular}

\end{ruledtabular}
\end{table*}

\begin{table*}
\caption{\label{tab:geos2}
Electronic energies (eV), rovibrational constants (cm$^{-1}$), and geometries (\AA, degrees) for group 15 MSH molecules considered in Section II. 
}
\begin{ruledtabular}
\begin{tabular}{lllllllllll}

PSH & $\Delta E$ & $r$(P-S) & $r$(S-H) & $\angle$(P-S-H) & A & B & C & $v_1$ & $v_2$ & $v_3$\\
\hline
$\tilde{X}^3A''$ & -- & 2.107 & 1.344 & 96.05 & 9.709 & 0.231 & 0.237 & 508.23 & 744.66 & 2693.56\\
$\tilde{a}^1A'$ & 0.325 & 1.991 & 1.359 & 105.02 & 10.157 & 0.256 & 0.263 & 600.19 & 889.94 & 2533.13\\
$\tilde{b}^1A''$ & 0.856 & 2.078 & 1.347 & 97.41 & 9.733 & 0.237 & 0.243 & 535.22 & 744.00 & 2659.74\\
$\tilde{A}^3A'$ & 3.468 & 2.223 & 1.362 & 107.95 & 10.432 & 0.206 & 0.210 & 413.65 & 710.99 & 2526.34\\
$\tilde{B}^3A''$ \footnote[1]{State is unbound and pre-dissociative. Geometry and energies correspond to saddlepoints on the potential energy surface.} & 3.740 & 2.139 & 1.353 & 102.89 & 10.025 & 0.228 & 0.223 & 457.26$i$ & 450.37 & 2565.86\\
$\tilde{c}^1A'$ & 3.962 & 2.309 & 1.343 & 92.52 & 9.600 & 0.198 & 0.194 & 356.35 & 683.95 & 2698.57\\
\hline
AsSH & $\Delta E$ & $r$(As-S) & $r$(S-H) & $\angle$(As-S-H) & A & B & C & $v_1$ & $v_2$ & $v_3$\\
\hline
$\tilde{X}^3A''$ & -- & 2.222 & 1.343 & 95.30 & 9.689 & 0.149 & 0.146 & 401.33 & 712.08 & 2703.94\\
$\tilde{a}^1A'$ & 0.475 & 2.114 & 1.352 & 103.57 & 10.090 & 0.163 & 0.160 & 495.79 & 716.90 & 2613.70\\
$\tilde{b}^1A''$ & 0.852 & 2.196 & 1.345 & 96.49 & 9.710 & 0.152 & 0.149 & 429.09 & 658.02 & 2662.37\\
$\tilde{A}^3A'$ & 3.193 & 2.518 & 1.342 & 86.22 & 9.605 & 0.116 & 0.115 & 249.63 & 457.25 & 2712.55\\
$\tilde{B}^3A''$ & 3.831 & 2.321 & 1.351 & 104.03 & 9.605 & 0.116 & 0.115 & 194.00 & 414.32 & 2727.01\\
$\tilde{c}^1A'$ & 3.788 & 2.508 & 1.341 & 86.59 & 9.607 & 0.117 & 0.116 & 247.95 & 638.28 & 2720.60\\
\hline
SbSH & $\Delta E$ & $r$(Sb-S) & $r$(S-H) & $\angle$(Sb-S-H) & A & B & C & $v_1$ & $v_2$ & $v_3$\\
\hline
$\tilde{X}^3A''$ & -- & 2.427 & 1.342 & 94.71 & 9.672 & 0.110 & 0.109 & 353.07 & 671.24 & 2711.20\\
$\tilde{a}^1A'$ & 0.593 & 2.346 &  1.345 & 101.03 & 9.972 & 0.117 & 0.116 & 391.32 & 746.03 & 2670.33\\
$\tilde{b}^1A''$ & 0.790 & 2.410 & 1.343 & 95.58 & 9.692 & 0.112 & 0.110 & 363.68 & 666.81 & 2701.20\\
$\tilde{A}^3A'$ & 2.805 & 2.691 & 1.343 & 88.61 & 9.572 & 0.0899 & 0.0890 & 245.33 & 423.23 & 2707.10\\
$\tilde{B}^3A''$ & 3.681 & 2.551 & 1.347 & 102.06 & 10.011 & 0.0993 & 0.0983 & 297.11 & 551.80 & 2660.40\\
$\tilde{c}^1A'$ & 3.461 &  2.469 & 1.420 & 179.94 & -- & 0.102 & -- & 244.68 & 380.65 & 1935.27\\
\hline
BiSH & $\Delta E$ & $r$(Bi-S) & $r$(S-H) & $\angle$(Bi-S-H) & A & B & C & $v_1$ & $v_2$ & $v_3$\\
\hline
$\tilde{X}^3A''$ & -- & 2.516 & 1.341 & 94.45 & 9.678 & 0.0931 & 0.0923 & 323.28 & 645.81 & 2717.71\\
$\tilde{a}^1A'$ & 0.658 & 2.440 & 1.343 & 100.18 & 9.944 & 0.0987 & 0.0978 & 354.46 & 706.33 & 2692.61\\
$\tilde{b}^1A''$ & 0.775 & 2.501 & 1.342 & 95.13 & 9.678 & 0.0931 & 0.0923 & 330.65 & 386.65 & 2705.49\\
$\tilde{A}^3A'$ & 2.583 & 2.774 & 1.342 & 88.69 & 9.580 & 0.0769 & 0.0763 & 231.47 & 334.45 & 2702.03\\
$\tilde{B}^3A''$ & 3.728 & 2.763 & 1.346 & 106.00 & 10.396 & 0.0768 & 0.0762 & 221.64 & 462.20 & 2678.71\\
$\tilde{c}^1A'$ & 3.111 & 2.309 &  1.341 & 87.77 & 9.590 & 0.0774 & 0.0767 & 232.26 & 379.37 & 2705.33\\
\end{tabular}

\end{ruledtabular}
\end{table*}

\begin{table*}
\caption{\label{tab:FCF_expanded}
Calculated Franck-Condon factors (FCFs) for eight leading vibronic decays for cycling transitions in multivalent molecules. Vibrational states are denoted with the shorthand convention $(v_1\,v_2\,v_3)$.
}
\begin{ruledtabular}
\begin{tabular}{llllllll}
\multicolumn{2}{c}{BSH} & \multicolumn{2}{c}{AlSH} & \multicolumn{2}{c}{GaSH} & \multicolumn{2}{c}{InSH}\\
\cline{1-2} \cline{3-4} \cline{5-6} \cline{7-8}
$\tilde{b}^3A''\to\tilde{X}^1A'$ & FCF & $\tilde{b}^3A''\to\tilde{X}^1A'$ & FCF & 
$\tilde{b}^3A''\to\tilde{X}^1A'$ & FCF &
$\tilde{b}^3A''\to\tilde{X}^1A'$ & FCF\\
\cline{1-2} \cline{3-4} \cline{5-6} \cline{7-8}
$(000)\to(000)$ & 0.9600 & $(000)\to(000)$ &  0.9974 & $(000)\to(000)$ & 0.9804 & $(000)\to(000)$ & 0.9517\\
$(000)\to(100)$ & $0.01497$ & $(000)\to(100)$ & $2.373\times 10^{-3}$ & $(000)\to(100)$ & 0.01139 & $(000)\to(100)$ & 0.03719\\
$(000)\to(010)$ & $0.01511$ & $(000)\to(010)$ & $2.309\times 10^{-4}$ & $(000)\to(020)$ & $5.910\times 10^{-3}$ & $(000)\to(020)$ & $5.854\times 10^{-3}$\\
$(000)\to(200)$ & $7.835\times 10^{-3}$ & $(000)\to(101)$ & $3.192\times 10^{-6}$ & $(000)\to(110)$ & $8.691\times 10^{-4}$ & $(000)\to(010)$ & $2.345\times10^{-3}$\\
$(000)\to(020)$ & $4.236\times 10^{-4}$ & $(000)\to(011)$ & $2.369\times 10^{-6}$ & $(000)\to(200)$ & $6.751\times 10^{-4}$ & $(000)\to(200)$ & $2.123\times10^{-3}$\\
$(000)\to(300)$ & $3.093\times 10^{-4}$ & $(000)\to(200)$ & $1.384\times 10^{-6}$ & $(000)\to(010)$ & $5.467\times 10^{-4}$ & $(000)\to(120)$ & $2.765\times10^{-4}$\\
$(000)\to(110)$ & $1.467\times 10^{-4}$ & $(000)\to(110)$ & $7.536\times 10^{-7}$ & $(000)\to(120)$ & $8.754\times 10^{-5}$ & $(000)\to(011)$ & $1.861\times10^{-4}$\\
$(000)\to(400)$ & $9.401\times 10^{-5}$ & $(000)\to(020)$ & $2.257\times 10^{-7}$ & $(000)\to(011)$ & $6.548\times 10^{-5}$ & $(000)\to(110)$ & $1.404\times10^{-4}$\\
\cline{1-2} \cline{3-4} \cline{5-6} \cline{7-8}
$\tilde{b}^3A''\to\tilde{a}^3A'$ & FCF & $\tilde{b}^3A''\to\tilde{a}^3A'$ & FCF & 
$\tilde{b}^3A''\to\tilde{a}^3A'$ & FCF &
$\tilde{b}^3A''\to\tilde{a}^3A'$ & FCF\\
\cline{1-2} \cline{3-4} \cline{5-6} \cline{7-8}
$(000)\to(010)$ & $0.2466$ & $(000)\to(000)$ & $0.5645$ & $(000)\to(000)$ & $0.3658$ & $(000)\to(000)$ & 0.5455\\
$(000)\to(000)$ & $0.1708$ & $(000)\to(010)$ & $0.2485$ & $(000)\to(100)$ & $0.2125$ & $(000)\to(100)$ & 0.2076\\
$(000)\to(020)$ & $4.980\times 10^{-3}$ & $(000)\to(020)$ & $0.06006$ & $(000)\to(010)$ & $0.1357$ & $(000)\to(010)$ & 0.1084\\ 
$(000)\to(510)$ & $6.612\times 10^{-3}$ & $(000)\to(110)$ & $0.01666$ & $(000)\to(200)$ & $0.07661$ & $(000)\to(200)$ & 0.05159\\ 
$(000)\to(120)$ & $6.115\times 10^{-3}$ & $(000)\to(030)$ & $0.01311$ & $(000)\to(110)$ & $0.06946$ & $(000)\to(110)$ & 0.03542\\
$(000)\to(101)$ & $5.978\times 10^{-3}$ & $(000)\to(101)$ & $0.01020$ & $(000)\to(020)$ & $0.03233$ & $(000)\to(020)$ & 0.01413\\
$(000)\to(700)$ & $5.719\times 10^{-3}$ & $(000)\to(111)$ & $5.778\times 10^{-3}$ & $(000)\to(210)$ & $0.02251$ & $(000)\to(300)$ & 0.01023\\
$(000)\to(020)$ & $4.980\times 10^{-3}$ & $(000)\to(120)$ & $2.694\times 10^{-3}$ & $(000)\to(300)$ & $0.02154$ & $(000)\to(210)$ & $7.788\times 10^{-3}$\\
\hline
\multicolumn{2}{c}{TlSH} & \multicolumn{2}{c}{CSH} & \multicolumn{2}{c}{SiSH} & \multicolumn{2}{c}{GeSH}\\
\cline{1-2} \cline{3-4} \cline{5-6} \cline{7-8}
$\tilde{b}^3A''\to\tilde{a}^3A'$ & FCF & $\tilde{a}^4A''\to\tilde{X}^2A'$ & FCF & $\tilde{a}^4A''\to\tilde{X}^2A'$ & FCF & $\tilde{a}^4A''\to\tilde{X}^2A'$ & FCF\\
\cline{1-2} \cline{3-4} \cline{5-6} \cline{7-8}
$(000)\to(000)$ & 0.5210 & $(000)\to(000)$ & $0.6667$ & $(000)\to(000)$ & $0.7049$ & $(000)\to(000)$ & $0.2498$\\
$(000)\to(100)$ & 0.2174 & $(000)\to(100)$ & $0.1855$ & $(000)\to(100)$ & $0.2081$ & $(000)\to(100)$ & $0.2508$\\
$(000)\to(200)$ & $0.1209$ & $(000)\to(010)$ & $0.07485$ & $(000)\to(200)$ & $0.05043$ & $(000)\to(200)$ & $0.1979$\\
$(000)\to(300)$ & 0.05302 & $(000)\to(110)$ & $0.03172$ & $(000)\to(010)$ & $0.01917$ & $(000)\to(300)$ & $0.1307$\\
$(000)\to(400)$ & 0.02369 & $(000)\to(200)$ & $0.01888$ & $(000)\to(300)$ & $0.01028$ & $(000)\to(400)$ & $0.07796$\\
$(000)\to(010)$ & 0.01794 & $(000)\to(020)$ & $6.588\times 10^{-3}$ & $(000)\to(110)$ & $2.614\times 10^{-3}$ & $(000)\to(500)$ &  $0.04304$\\
$(000)\to(110)$ & 0.01231 & $(000)\to(210)$ & $5.073\times 10^{-3}$ & $(000)\to(400)$ & $1.909\times 10^{-3}$ & $(000)\to(600)$ & $0.02243$\\
$(000)\to(500)$ & $9.774\times 10^{-3}$ & $(000)\to(120)$ & $3.459\times 10^{-3}$ & $(000)\to(011)$ & $1.174\times 10^{-3}$ & $(000)\to(700)$ & $0.01116$\\
\cline{1-2} \cline{3-4} \cline{5-6} \cline{7-8}
$\tilde{b}^3A''\to\tilde{a}^3A'$ & FCF & $\tilde{a}^4A''\to\tilde{X}^2A'$ & FCF & $\tilde{a}^4A''\to\tilde{X}^2A'$ & FCF & $\tilde{a}^4A''\to\tilde{X}^2A'$ & FCF\\
\cline{1-2} \cline{3-4} \cline{5-6} \cline{7-8}
$(000)\to(000)$ & 0.2706 & $(000)\to(000)$ & $0.3412$ & $(000)\to(000)$ & $0.7515$ & $(000)\to(000)$ & $0.5631$\\
$(000)\to(100)$ & 0.2362 & $(000)\to(100)$ & $0.2235$ & $(000)\to(010)$ & $0.2083$ & $(000)\to(100)$ & $0.1894$\\
$(000)\to(200)$ & 0.1407 & $(000)\to(010)$ & $0.1691$ & $(000)\to(100)$ & $0.01423$ & $(000)\to(010)$ &  $0.08689$\\
$(000)\to(010)$ & 0.07526 & $(000)\to(110)$ & $0.1067$ & $(000)\to(020)$ & $0.01265$ & $(000)\to(200)$  & $0.08541$\\
$(000)\to(300)$ & 0.06789 & $(000)\to(020)$ & $0.03838$ & $(000)\to(011)$ & $5.485\times 10^{-3}$ & $(000)\to(300)$  & $0.03029$\\
$(000)\to(110)$ & 0.05896 & $(000)\to(120)$ & $0.02814$ & $(000)\to(120)$ & $2.550\times 10^{-3}$ & $(000)\to(300)$  & $0.03029$\\
$(000)\to(210)$ & 0.03404 & $(000)\to(120)$ & $0.02323$ & $(000)\to(021)$ & $2.135\times 10^{-3}$ & $(000)\to(110)$  & $0.01331$\\
$(000)\to(400)$ & 0.02870 & $(000)\to(111)$ & $7.765\times 10^{-3}$ & $(000)\to(001)$ & $1.159\times 10^{-3}$ & $(000)\to(400)$  & $0.01097$\\
\hline
\multicolumn{2}{c}{PSH} & \multicolumn{2}{c}{AsSH} & \multicolumn{2}{c}{SbSH} & \multicolumn{2}{c}{BiSH}\\
\cline{1-2} \cline{3-4} \cline{5-6} \cline{7-8}
$\tilde{b}^1A''\to\tilde{X}^3A'$ & FCF & $\tilde{b}^1A''\to\tilde{X}^3A'$ & FCF & $\tilde{b}^1A''\to\tilde{X}^3A'$ & FCF & $\tilde{b}^1A''\to\tilde{X}^3A'$ & FCF\\
\cline{1-2} \cline{3-4} \cline{5-6} \cline{7-8}
$(000)\to(000)$ & $0.9018$ & $(000)\to(000)$ & $0.9224$ & $(000)\to(000)$ & $0.9572$ & $(000)\to(000)$ & $0.9674$\\
$(000)\to(100)$ & $9.074\times 10^{-2}$ & $(100)\to(000)$ & $0.06913$ & $(000)\to(100)$ & $0.03952$ & $(000)\to(020)$ & $0.03050$\\
$(000)\to(010)$ & $3.852\times 10^{-3}$ & $(000)\to(010)$ & $5.412\times 10^{-3}$ & $(000)\to(010)$ & $2.684\times10^{-3}$ & $(000)\to(040)$ & $1.443\times 10^{-3}$\\
$(000)\to(200)$ & $2.506\times 10^{-3}$ & $(000)\to(020)$ & $1.362\times 10^{-3}$ & $(000)\to(200)$ & $3.351\times 10^{-4}$ & $(000)\to(011)$ & $4.409\times 10^{-4}$\\
$(000)\to(110)$ & $6.621\times 10^{-4}$ & $(000)\to(110)$ & $5.081\times 10^{-4}$ & $(000)\to(110)$ & $2.042\times 10^{-4}$ & $(000)\to(060)$ & $7.583\times 10^{-5}$\\
$(000)\to(001)$ & $2.229\times 10^{-4}$ & $(000)\to(200)$ & $4.726\times 10^{-4}$ & $(000)\to(011)$ & $7.613\times 10^{-5}$ & $(000)\to(200)$ & $7.041\times 10^{-5}$\\
$(000)\to(011)$ & $1.648\times 10^{-4}$ & $(000)\to(011)$ & $2.257\times 10^{-4}$ & $(000)\to(001)$ & $2.977\times 10^{-5}$ & $(000)\to(110)$ & $4.556\times 10^{-5}$\\
$(000)\to(101)$ & $4.968\times 10^{-5}$ & $(000)\to(210)$ & $2.208\times 10^{-4}$ & $(000)\to(020)$ & $1.709\times 10^{-5}$ & $(000)\to(031)$ & $4.171\times 10^{-5}$ \\
\cline{1-2} \cline{3-4} \cline{5-6} \cline{7-8}
$\tilde{b}^1A''\to\tilde{a}^1A'$ & FCF & $\tilde{b}^1A''\to\tilde{a}^1A'$ & FCF & $\tilde{b}^1A''\to\tilde{a}^1A'$ & FCF & $\tilde{b}^1A''\to\tilde{a}^1A'$ & FCF\\
\cline{1-2} \cline{3-4} \cline{5-6} \cline{7-8}
$(000)\to(000)$ & $0.2986$ & $(000)\to(000)$ & $0.2690$ & $(000)\to(000)$ & $0.5130$ & $(000)\to(000)$ & $0.3994$\\
$(000)\to(100)$ & $0.2954$ & $(000)\to(100)$ & $0.2780$ & $(000)\to(100)$ & $0.2797$ & $(000)\to(100)$ & $0.3044$\\
$(000)\to(200)$ & $0.1633$ & $(000)\to(200)$ & $0.1616$ & $(000)\to(200)$ & $0.08656$ & $(000)\to(200)$ & $0.1267$\\
$(000)\to(300)$ & $0.06617$ & $(000)\to(300)$ & $0.06906$ & $(000)\to(010)$ & $0.05086$ & $(000)\to(300)$ & $0.03802$ \\
$(000)\to(010)$ & $0.04189$ & $(000)\to(010)$ & $0.06129$ & $(000)\to(110)$ & $0.02369$ & $(000)\to(020)$ & $0.02171$\\
$(000)\to(110)$ & $0.03943$ & $(000)\to(110)$ & $0.05213$ & $(000)\to(300)$ & $0.01983$ & $(000)\to(010)$ & $0.02093$\\
$(000)\to(400)$ & $0.02183$ & $(000)\to(210)$ & $0.02505$ & $(000)\to(210)$ & $6.316\times 10^{-3}$ & $(000)\to(120)$ & $0.01669$ \\
$(000)\to(210)$ & $0.02083$ & $(000)\to(400)$ & $0.02412$ & $(000)\to(020)$ & $5.834 \times 10^{-3}$ & $(000)\to(110)$ & $0.01638$\\
\end{tabular}
\end{ruledtabular}
\end{table*}

\begin{table*}
\caption{\label{tab:SOC}
Computed mean-field Breit-Pauli spin-orbit matrix elements (cm$^{-1}$) for group 13 MSH molecules with respect to spin-electronic states $|\Lambda, \Sigma\ket$.
}
\begin{ruledtabular}
\begin{tabular}{cccccc}
Matrix Element & BSH & AlSH & GaSH & InSH & TlSH\\
\hline
$\bra \tilde{X}^1A',0|\hat{H}_\text{SO}|\tilde{b}^3A'',\pm 1\ket$ & $-34.041e^{\mp 0.964i}$ & $43.203e^{\pm 1.333i}$ & $172.641e^{\pm 3.563i}$ & $518.827e^{\pm 9.871i}$ & $209.677e^{\mp 6140.860i}$\\
$\bra \tilde{A}^1A',0|\hat{H}_\text{SO}|\tilde{b}^3A'',\pm 1\ket$ & $-5.078e^{\mp 6.332i}$ & $2.821e^{\mp 28.488i}$ & $-4.5965e^{\pm 202.794i}$ & $25.886e^{\mp 793.378i}$ & $5212.156e^{\pm 332.300i}$\\
$\bra \tilde{B}^1A'',0|\hat{H}_\text{SO}|\tilde{a}^3A',\pm 1\ket$ & $-2.485e^{\pm 8.770i}$ & $0.0863e^{\mp 29.389i}$ & $10.439e^{\mp 205.505i}$ & $32.190e^{\mp 806.937i}$ & $-5123.191e^{\mp 216.603i}$\\
$\bra \tilde{C}^1A',0|\hat{H}_\text{SO}|\tilde{b}^3A',\pm 1\ket$ & $7.992e^{\mp 30.267i}$ & $5.388e^{\pm 16.171i}$ & $23.325e^{\mp 9.280i}$ & $257.066e^{\pm 31.535i}$ & $-5206.329e^{\mp 540.624i}$\\
\hline
& CSH & SiSH & GeSH \\
\hline
$\bra \tilde{X}^2A',\pm \frac{1}{2}|\hat{H}_\text{SO}|\tilde{a}^4A'',\pm \frac{3}{2}\ket$ & $51.645e^{\pm 8.795i}$ & $-71.321e^{\mp 5.719i}$ & $64.008e^{\pm 6.129i}$\\
$\bra \tilde{X}^2A',\pm \frac{1}{2}|\hat{H}_\text{SO}|\tilde{a}^4A'',\mp \frac{1}{2}\ket$ & $29.817e^{\mp 5.078i}$ & $-41.177e^{\mp 3.302i}$ & $36.955e^{\mp 3.538i}$\\
$\bra \tilde{B}^2A',\pm \frac{1}{2}|\hat{H}_\text{SO}|\tilde{a}^4A'',\pm \frac{3}{2}\ket$ & $15.329e^{\pm 4.737i}$ & $12.027e^{\mp 60.937i}$ & $13.746e^{\pm 32.229i}$ \\
$\bra \tilde{B}^2A',\pm \frac{1}{2}|\hat{H}_\text{SO}|\tilde{a}^4A'',\mp \frac{1}{2}\ket$ & $8.850e^{\mp 2.735i}$ & $6.944e^{\pm 35.182i}$ & $7.936e^{\mp 18.608i}$\\
$\bra \tilde{C}^2A',\pm \frac{1}{2})|\hat{H}_\text{SO}|\tilde{a}^4A'',\pm \frac{3}{2})\ket$ & $4.635e^{-20.198i}$ & $80.457e^{-12.007i}$ & $93.260e^{12.370i}$
\\
$\bra \tilde{C}^2A',\pm \frac{1}{2}|\hat{H}_\text{SO}|\tilde{a}^4A'',\mp \frac{1}{2}\ket$ & $86.181e^{23.323}$ & $92.904e^{-13.864i}$ & $-107.687e^{14.284i}$\\
\hline
& PSH & AsSH & SbSH & BiSH \\
\hline
$\bra \tilde{X}^3A'',\pm 1|\hat{H}_\text{SO}|\tilde{a}^1A',0\ket$ & $-2.608e^{\mp 103.992i}$ & $12.889e^{\mp592.464i}$ & $-42.417e^{\pm 1880.700i}$ & $10272.392e^{\pm 247.815i}$\\
$\bra \tilde{A}^3A', \pm 1|\hat{H}_\text{SO}|\tilde{b}^1A'', 0 \ket$ & $16.024e^{\pm 13.687i}$ & $372.723e^{\pm 37.734i}$ & $1289.094e^{\pm 52.616i}$ & $-368.162e^{\mp 8585.580i}$\\
$\bra \tilde{B}^3A'',\pm 1|\hat{H}_\text{SO}|\tilde{a}^1A',0\ket$ & $117.880e^{\pm 9.027i}$ & $-428.892e^{\pm 4.021i}$ & $1162.687e^{\pm 0.326}$ & $-7077.983e^{\pm 166.707i}$ \\
$\bra \tilde{C}^3A', \pm 1|\hat{H}_\text{SO}|\tilde{b}^1A'',0\ket$ & -- & $-16.942e^{\pm 114.620i}$ & $902.415e^{\pm 31.278i}$ & --
\end{tabular}
\end{ruledtabular}
\end{table*}

\begin{table*}
\caption{\label{tab:tdm}
Calculated moments (in Debye) for spin-allowed dipole transitions in multivalent MSH molecules
}
\begin{ruledtabular}
\begin{tabular}{lllllllll}
Transition & Orientation & BSH & AlSH & GaSH & InSH & TlSH\\
\hline
$d_0(\tilde{X}^1A')$ & $ab$-type & 1.7644 & 1.3585 & 1.8616 & 2.7739 & 3.7892\\
$\tilde{X}^1A'\to \tilde{A}^1A'$ & $ab$-type & 1.5059 & 2.5257 & 1.9927 & 2.1981 & 1.4057\\
$\tilde{X}^1A'\to \tilde{B}^1A''$ & $c$-type & 1.5052 & 2.6114 & 1.9376 & 2.1372 & 1.3434\\
$\tilde{X}^1A'\to \tilde{C}^1A'$ & $ab$-type & 1.5051 &  0.3171 & 0.9196 & 0.7434 & 0.9694\\
$\tilde{a}^3A'\to \tilde{b}^3A''$ & $c$-type & 0.06753 & 0.02584 & 0.029131 & 0.01351 & 0.005602\\
\hline
& & CSH & SiSH & GeSH\\
\hline
$d_0(\tilde{X}^2A')$ & $ab$-type & 1.4663 & 0.8947 & 1.3986\\
$\tilde{X}^2A'\to \tilde{A}^2A''$ & $c$-type & 0.1008 & 0.02842 & 0.02842\\
$\tilde{A}^2A''\to \tilde{B}^2A'$ & $c$-type & 0.02508 & 0.2094 & 0.2551\\
$\tilde{A}^2A''\to \tilde{C}^2A'$ & $c$-type & 0.1474 & 0.07286 & 0.03832\\
\hline
& & PSH & AsSH & SbSH & BiSH\\
\hline
$d_0(\tilde{X}^3A'')$ & $ab$-type & 0.8065 & 0.9317 & 1.3588 & 2.0861\\
$\tilde{X}^3A''\to \tilde{A}^3A'$ & $c$-type & 0.30537 & 0.3776 & 0.1753 & 0.4578\\
$\tilde{X}^3A''\to \tilde{B}^3A''$ & $ab$-type & 0.67851 & 0.6206 &  0.6950 & 0.4795\\
$\tilde{X}^3A''\to \tilde{C}^3A'$ & $c$-type & 0.29843 & 0.07096 & 0.0712 & 0.06363\\
$\tilde{a}^1A'\to \tilde{b}^3A''$ & $c$-type & 0.06075 & 0.05008 & 0.03274 & 0.02700\\
\end{tabular}
\end{ruledtabular}
\end{table*}
\FloatBarrier
\end{document}